\documentclass[12pt,english]{article}
\usepackage[T1]{fontenc}
\usepackage[utf8]{inputenc}   
\usepackage{microtype}
\usepackage{comment}
\usepackage{tikz}
\usetikzlibrary{arrows.meta,positioning,calc,patterns,decorations.pathreplacing,fit,backgrounds}
\usepackage{float}
\usepackage{graphicx} 
\usepackage{amsmath}
\usepackage{amsthm}
\usepackage{amssymb}
\usepackage{bm}
\usepackage{mathtools}
\usepackage{newtxtext}
\usepackage[cmintegrals,varg]{newtxmath}

\theoremstyle{plain} 

\newtheorem{Definition}{Definition}

\theoremstyle{remark} 

\usepackage[authoryear,longnamesfirst]{natbib}
\setcitestyle{round}

\usepackage{tabularx}
\usepackage{siunitx}
\usepackage{booktabs}
\usepackage{array}
\usepackage{longtable}
\usepackage{multirow}
\usepackage{threeparttable}
\usepackage[table]{xcolor}
\usepackage{ragged2e}
\usepackage{lscape}
\usepackage{rotating}
\usepackage{rotfloat}
\usepackage{placeins}
\usepackage{caption}

\newcolumntype{C}[1]{>{\centering\arraybackslash}p{#1}}
\newcolumntype{J}[1]{>{\justify\arraybackslash}p{#1}}
\newcolumntype{R}[1]{>{\RaggedLeft\arraybackslash}p{#1}}
\newcolumntype{L}[1]{>{\RaggedRight\arraybackslash}p{#1}}
\newcolumntype{G}{@{\extracolsep{0.5cm}}l@{\extracolsep{0pt}}}%
\newcolumntype{P}[1]{>{\centering\arraybackslash}p{#1}}
\newcolumntype{Y}{>{\centering\arraybackslash}X}

\newcommand{\nhphantom}[1]{\sbox0{#1}\hspace{-\the\wd0}} 

\definecolor{mypink}{rgb}{0.961,0.839,0.929}
\definecolor{Gray}{gray}{0.9}

\everymath{\setlength{\abovedisplayskip}{10pt}\setlength{\belowdisplayskip}{10pt}}
\everydisplay{\setlength{\abovedisplayskip}{10pt}\setlength{\belowdisplayskip}{10pt}}

\usepackage[verbose,tmargin=2cm,bmargin=2.2cm,lmargin=2.2cm,rmargin=2.2cm]{geometry}
\usepackage{setspace}
\usepackage[colorlinks=true,linkcolor=blue,citecolor=blue,urlcolor=blue]{hyperref}

\title{Split-Session Cluster GARCH for Overnight and Intraday Returns: The Role of Tail Heterogeneity}

\author{Xinxian Chen$^{\ddagger}$ \qquad Peter Reinhard Hansen$^{\mathsection}$\thanks{Corresponding author: Peter Reinhard Hansen (hansen@unc.edu). Chen Tong acknowledges financial support from the Youth Fund of the National Natural Science Foundation of China (72301227) and the Fujian Provincial Natural Science Foundation of China (2025J08008).}\qquad Chen Tong$^{\ddagger}$
\\[0.2cm] \small $^{\ddagger}$Department of Finance, School of Economics, Xiamen University, China
\\[-0.1cm] \small $^{\mathsection}$Department of Economics, University of North Carolina at Chapel Hill, United States}

\date{\today}

\begin{document}
\maketitle

\vspace{-0.8cm}

\begin{abstract}
We propose the Split-Session Cluster GARCH model for heavy-tailed multivariate dependence among asset returns decomposed into overnight and intraday components. The model uses convolution-$t$ distributions to allow tail behavior to differ across clusters defined by trading sessions and, within each session, by economic sectors. It also accommodates block-structured conditional correlation matrices, preserving parsimony and scalability in high-dimensional settings. The resulting likelihood remains tractable and yields a score-driven specification for dynamic correlations. We apply the model to U.S. equity returns in six-asset and 100-asset applications. The results reveal pronounced tail heterogeneity between overnight and intraday returns. Model comparisons show that session-specific tail parameters substantially improve fit relative to a common multivariate-$t$ specification, while sector-level tail partitioning delivers additional gains concentrated mainly in the overnight component. In the 100-asset application, asset-level tail heterogeneity delivers the strongest out-of-sample likelihood and global minimum-variance (GMV) portfolio performance.
\end{abstract}


\bigskip

\section{Introduction}
Understanding dependence among financial asset returns is central to portfolio
allocation, risk management, and asset pricing. A large literature models
conditional covariance matrices by separating volatility and correlation
dynamics, as in the CCC model of \citet{Bollerslev1990} and the DCC model of
\citet{Engle2002}; see also \citet{Tse2002}, \citet{Aielli2013}, and
\citet{Engle2019}. Recent studies further develop dynamic covariance and
correlation forecasts for equity markets, portfolio selection, and systemic-risk measurement \citep{Symitsi2018, Moura2020,
DeNard2022, HonigKircher2025, GirardiErgun2013}. Despite these advances,
dynamic correlation modeling remains challenging because the number of
correlations grows quadratically with the number of assets, positive
definiteness must be preserved, and cross-product-based updates can be sensitive
to extremes. These issues are amplified in heavy-tailed multivariate systems,
where misspecified tail behavior can distort dynamic dependence updates.

To address these challenges, a growing literature imposes structured
representations on large correlation systems. One strand restricts correlation
dynamics through factor structures (e.g., \citealp{Creal2015};
\citealp{Oh2023a}), while another uses block correlation matrices
(e.g., \citealp{Engle2012}; \citealp{Tong2026}; \citealp{Archakov2026}) or
combines factor and block structures \citep{TongHansen2026}. In parallel, the
correlation parameterization of \citet{Archakov2021} maps unconstrained
real-valued parameters into positive definite correlation matrices. Separately,
the canonical block representation of \citet{Archakov2024} exploits block
structure to reduce the dimension of the correlation system from the asset-pair
level to the block-pair level. The score-driven framework of \citet{Creal2013}
provides a likelihood-based method for updating time-varying parameters in
dynamic dependence models.

Most multivariate volatility and correlation studies use daily close-to-close
returns, although these returns combine overnight close-to-open and intraday
open-to-close components with distinct information-arrival mechanisms. During
trading hours, prices adjust continuously through trading, whereas overnight
information accumulates and is incorporated when the market reopens. Consistent
with this distinction, \citet{French1986} and \citet{LockwoodMcInish1990}
document sharp differences between trading and non-trading returns, while
\citet{MoshirianNguyenPham2012}, \citet{Barclay2003}, and \citet{Lou2019}
show that overnight information, after-hours trading, and heterogeneous investor
clienteles affect price discovery and return dynamics. These findings suggest
that treating daily returns as homogeneous close-to-close objects may obscure
important session-level differences in distributions and cross-asset
dependence.

A more recent literature explicitly models overnight and intraday returns
separately. \citet{Blanc2014a} document important differences in the volatility dynamics of the two trading sessions. \citet{Linton2020} propose a
semiparametric coupled component DCS-EGARCH model for intraday and overnight
volatility. \citet{Dhaene2020} develop mixed-frequency multivariate GARCH
models that combine high-frequency intraday returns and overnight returns to forecast lower-frequency covariance matrices.  \citet{KangBabbs2012} introduce a multivariate copula-GARCH model for
overnight and intraday returns with DCC-type dependence. Relatedly, based on high-frequency intraday data, \citet{Kim2023} propose a univariate overnight GARCH-It\^{o} model with separate open-to-close and close-to-open volatility processes, and
\citet{Kim2024} extend this idea to large volatility matrix estimation and
prediction in a high-dimensional factor framework.

While these studies provide important insights into session-level volatility
and dependence modeling, several rely on high-frequency intraday data or
realized measures. By contrast, we model the dynamic correlation structure of
multivariate overnight and intraday return innovations using only daily open and close
prices, which are widely available across assets and markets. More importantly,
existing cross-asset correlation models often pool overnight and intraday
components and typically impose a common tail structure across assets or broad
asset groups. Such restrictions are problematic when tail behavior differs
across both trading sessions and economically meaningful groups such as sectors. A flexible multivariate heavy-tailed model should therefore allow tail
heterogeneity across both time and cross-sectional dimensions while preserving a
tractable and positive definite dynamic correlation structure.

In this paper, we develop the Split-Session Cluster GARCH model for
heavy-tailed multivariate dependence among overnight and intraday return
innovations. The model extends the Cluster GARCH framework of \citet{Tong2026},
which was based on daily close-to-close returns, to a setting in which overnight
and intraday returns are modeled as distinct but related components. The term
``Split-Session'' reflects the decomposition of daily returns into overnight and
intraday components, while ``Cluster'' refers to the partitioning of innovations
into economically meaningful groups, such as trading sessions and sectors, each
with its own tail parameter under the convolution-$t$ distribution of \citet{Hansen2026}. The proposed
framework combines this clustered tail specification with the unconstrained
correlation parameterization of \citet{Archakov2021}, ensuring positive
definiteness of the dynamic correlation matrix. To improve scalability, we
further introduce block correlation structures for the overnight and intraday
correlation matrices and use their canonical representation
\citep{Archakov2024}.

This paper makes four main contributions. First, we extend the overnight--intraday return literature from volatility modeling to dynamic multivariate dependence modeling. Our approach explicitly models the conditional correlation matrices of standardized overnight and intraday innovations and allows correlation updates to differ across trading sessions. Second, we use the convolution-$t$ distributional framework of \citet{Hansen2026} to study session- and group-specific tail heterogeneity in a dynamic correlation setting. In our application, clusters are defined by trading sessions and, within each session, by economic groups such as sectors. This allows the model to distinguish between common-tail multivariate-$t$ specifications, session-specific tail specifications, sector-level tail partitions, and fully asset-specific Hetero-$t$ specifications. The empirical results show that tail heterogeneity is not merely a marginal distributional feature: the distinction between overnight and intraday tails materially affects score-driven correlation updating and model fit. Third, we provide a scalable implementation of the split-session dependence model by combining score-driven correlation dynamics with the canonical block-correlation representation. This yields a scalable dependence model in which correlation dynamics are updated at the block-pair rather than asset-pair level, reducing dimensionality while preserving positive definiteness. Fourth, we provide empirical evidence from both a six-asset application and a 100-asset application. The six-asset application, based on 20 years of daily stock returns for representative U.S. equities from two industrial sectors, allows us to examine the mechanisms of session-level and sector-level tail heterogeneity in detail. The 100-asset application shows that the block specification remains tractable in larger cross sections and that the Hetero-$t$ model performs best, suggesting that asset-level tail heterogeneity becomes more valuable as the dimension grows. We further assess the economic relevance of the high-dimensional forecasts using global minimum variance (GMV) portfolio performance.

The paper proceeds as follows. Section \ref{sec:The-univariate-extended}
introduces the asset-level Coupled EGARCH model for overnight and intraday volatility. Section \ref{sec:Dynamic-Correlations-and} presents the dynamic correlation
matrix framework. Section \ref{sec:Convolution-t-Distributions} introduces the
multivariate-$t$, Gaussian, Cluster-$t$, and      Hetero-$t$ specifications as special cases of the convolution-$t$ framework. Section
\ref{subsec:Scores-and-Fisher} derives the corresponding scores and Fisher information matrices, including the canonical block representation. 
Section \ref{sec:Benchmark-DCC} introduces the DCC benchmark model. 
Section
\ref{sec:Six-Asset-Application} presents the six-asset empirical application, while Section \ref{sec:High-Dimensional-Application} examines the 100-asset high-dimensional application. Section \ref{sec:Conclusion} concludes.

\section{The Coupled EGARCH Model\label{sec:The-univariate-extended}}

We decompose the daily close-to-close return of asset $i$ on day $t$, $R_{i,t}$, into an overnight return,
$R_{i,t}^{N}$, and an intraday return, $R_{i,t}^{D}$, where
$R_{i,t}^{N}=\log(P_{i,t}^{O}/P_{i,t-1}^{C})$ and
$R_{i,t}^{D}=\log(P_{i,t}^{C}/P_{i,t}^{O})$.
Following \citet{Linton2020}, the timeline is:

\begin{tikzpicture}[
	every node/.style = {font=\small},
	arr/.style = {-{Stealth}, thick}
	]
	
	\node (dots_l)  {$\cdots$};
	\node (Sc_prev)[right=0.8cm of dots_l]  {$P_{i,t-1}^{C}$};

	\coordinate (V1) at ([xshift=0.8cm]Sc_prev.east);

	\coordinate (N1_s) at ([xshift=0.6cm]V1);
	\coordinate (N1_e) at ([xshift=1.8cm]N1_s); 

	\node[anchor=west, inner sep=2pt] (So) at ([xshift=0.2cm]N1_e) {$P_{i,t}^{O}$};

	\coordinate (D1_s) at ([xshift=0.2cm]So.east);
	\coordinate (D1_e) at ([xshift=1.8cm]D1_s); 

	\node[anchor=west, inner sep=2pt] (Sc) at ([xshift=0.2cm]D1_e) {$P_{i,t}^{C}$};

	\coordinate (V2) at ([xshift=0.6cm]Sc.east);

	\coordinate (N2_s) at ([xshift=0.6cm]V2);
	\coordinate (N2_e) at ([xshift=1.8cm]N2_s);

	\node[anchor=west, inner sep=2pt] (So_next) at ([xshift=0.2cm]N2_e) {$P_{i,t+1}^{O}$};

	\node (dots_r)[right=0.8cm of So_next.east] {$\cdots$};

	\draw[arr] ([xshift=0.1cm]dots_l.east) -- ([xshift=-0.1cm]Sc_prev.west);
	\draw[arr] (N1_s) -- (N1_e) node[midway, above=0.05cm] {Night $t$};
	\draw[arr] (D1_s) -- (D1_e) node[midway, above=0.05cm] {Day $t$};
	\draw[arr] (N2_s) -- (N2_e) node[midway, above=0.05cm] {Night $t+1$};
	\draw[arr] ([xshift=0.1cm]So_next.east) -- ([xshift=-0.1cm]dots_r.west);

	\draw[thick] ([yshift=1.0cm]V1) -- ([yshift=-1.8cm]V1);
	\draw[thick] ([yshift=1.0cm]V2) -- ([yshift=-1.8cm]V2);

	\draw[decorate, decoration={brace, amplitude=5pt, mirror}, thick]
	([yshift=-0.2cm]N1_s) -- ([yshift=-0.2cm]N1_e)
	node[midway, below=6pt] {$R_{i,t}^{N}$};
	
	\draw[decorate, decoration={brace, amplitude=5pt, mirror}, thick]
	([yshift=-0.2cm]D1_s) -- ([yshift=-0.2cm]D1_e)
	node[midway, below=6pt] {$R_{i,t}^{D}$};
	
	\draw[decorate, decoration={brace, amplitude=5pt, mirror}, thick]
	([yshift=-0.2cm]N2_s) -- ([yshift=-0.2cm]N2_e)
	node[midway, below=6pt] {$R_{i,t+1}^{N}$};
	
    \draw[decorate, decoration={brace, amplitude=8pt, mirror}, thick]
    ([xshift=0.6cm, yshift=-1.1cm]V1) -- ([xshift=-0.6cm, yshift=-1.1cm]V2)
    node[midway, below=10pt] {$R_{i,t}$};
	
\end{tikzpicture}

 To capture the sequential nature of information within a trading day,
we define two nested information sets for asset $i$: $\mathcal{F}_{i,t}^{N}
=\sigma(\{R_{i,\tau}^{N},R_{i,\tau}^{D}\}_{\tau\leq t-1})$
and
$\mathcal{F}_{i,t}^{D}
=\sigma(\mathcal{F}_{i,t}^{N}\cup\{R_{i,t}^{N}\})$. Here, $\mathcal{F}_{i,t}^{N}$
contains all past overnight and intraday returns available at the close of day $t-1$, while $\mathcal{F}_{i,t}^{D}$ augments $\mathcal{F}_{i,t}^{N}$
by incorporating newly realized overnight return on day $t$.

We model the joint dynamics of overnight and intraday returns using
a vector autoregressive (VAR) framework. The conditional mean is given
by:
\begin{equation}
\left(\begin{array}{cc}
1 & 0\\
-\delta_i & 1
\end{array}\right)\left(\begin{array}{c}
R_{i,t}^{N}\\
R_{i,t}^{D}
\end{array}\right)=\left(\begin{array}{c}
\mu_{i}^{N}\\
\mu_{i}^{D}
\end{array}\right)+\left(\begin{array}{cc}
\phi_{11,i} & \phi_{12,i}\\
\phi_{21,i} & \phi_{22,i}
\end{array}\right)\left(\begin{array}{c}
R_{i,t-1}^{N}\\
R_{i,t-1}^{D}
\end{array}\right)+\left(\begin{array}{cc}
\sqrt{h_{i,t}^{N}} & 0\\
0 & \sqrt{h_{i,t}^{D}}
\end{array}\right)\left(\begin{array}{c}
Z_{i,t}^{N}\\
Z_{i,t}^{D}
\end{array}\right),\label{eq:EQ1}
\end{equation}

\noindent where $h_{i,t}^{N}=\operatorname{var}(R_{i,t}^{N}\mid\mathcal{F}_{i,t}^{N})$
and $h_{i,t}^{D}=\operatorname{var}(R_{i,t}^{D}\mid\mathcal{F}_{i,t}^{D})$
denote the conditional variances. The terms $Z_{i,t}^{N}$ and $Z_{i,t}^{D}$
are the standardized innovations with zero mean and unit variance.

Regarding the mean dynamics, the parameter $\delta_i$ captures the
contemporaneous impact of the overnight return on the intraday return
within the same trading day. Due to the triangular structure of the
system, the intraday return $R_{i,t}^{D}$ depends on the term $\delta_i R_{i,t}^{N}$.
As a result, a positive $\delta_i$ implies an overnight-intraday continuation, whereas
a negative $\delta_i$ indicates reversal. The vector $\left(\mu_{i}^{N},\mu_{i}^{D}\right)^{\prime}$
contains the intercepts of the two return components.
The matrix $\Phi_i=\left(\begin{smallmatrix}\phi_{11,i} & \phi_{12,i}\\
\phi_{21,i} & \phi_{22,i}
\end{smallmatrix}\right)$ captures both own-lag and cross-lag dependence in returns for asset $i$, allowing
past overnight and intraday returns to influence current returns across
trading sessions.

To capture asymmetric leverage effects and bidirectional volatility
spillovers between the overnight and intraday sessions, we model the
conditional log-variances using a Coupled Exponential GARCH (Coupled
EGARCH) specification. It is given by:
\begin{align*}
\log h_{i,t}^{N} & =\omega_{i}^{N}+\beta_{i}^{N}\log h_{i,t-1}^{N}+\tau_{1,i}^{N}Z_{i,t-1}^{N}+\tau_{2,i}^{N}\left|Z_{i,t-1}^{N}\right|+\underbrace{\delta_{1,i}^{N}Z_{i,t-1}^{D}+\delta_{2,i}^{N}\lvert Z_{i,t-1}^{D}\rvert}_{\text{spillover from previous day}},\nonumber \\
\log h_{i,t}^{D} & =\omega_{i}^{D}+\beta_{i}^{D}\log h_{i,t-1}^{D}+\tau_{1,i}^{D}Z_{i,t-1}^{D}+\tau_{2,i}^{D}\left|Z_{i,t-1}^{D}\right|+\underbrace{\delta_{1,i}^{D}Z_{i,t}^{N}+\delta_{2,i}^{D}\left|Z_{i,t}^{N}\right|}_{\text{spillover from current night}}.
\end{align*}

The EGARCH specification ensures positivity of conditional variances
without parameter restrictions and accommodates asymmetric responses
to positive and negative shocks. The intercepts $\omega_i^N$ and $\omega_i^D$, together with the persistence
parameters and average shock-magnitude terms, determine the baseline levels of
overnight and intraday log-volatility, respectively. The parameters
$\beta_i^N$ and $\beta_i^D$ measure the persistence in overnight and intraday
volatility. The coefficients \(\tau_{1,i}^c\) capture the sign effects, while
\(\tau_{2,i}^c\) capture the magnitude effects, for \(c\in\{N,D\}\). This structure allows volatility to respond asymmetrically
to positive and negative return innovations, reflecting the leverage
effect commonly observed in equity markets. The parameters $\delta_{1,i}^{N}$,
$\delta_{2,i}^{N}$, $\delta_{1,i}^{D}$, $\delta_{2,i}^{D}$ capture
volatility spillovers between overnight and intraday periods. Specifically,
$\delta_{1,i}^{N}$ and $\delta_{2,i}^{N}$ measure the impact of
intraday shocks from day $t-1$ on overnight volatility, while $\delta_{1,i}^{D}$
and $\delta_{2,i}^{D}$ capture the effect of overnight shocks on
intraday volatility within the same trading day.

We estimate the model asset by asset using quasi-maximum likelihood
estimation (QMLE) under the Gaussian assumption. The remaining distributional features of the standardized innovations,
including heavy tails and tail heterogeneity across sessions, are modeled in
the second-stage correlation model. The resulting standardized innovations,
$Z_{i,t}^{N}$ and $Z_{i,t}^{D}$, serve as the inputs to the multivariate
correlation model introduced in the next section, where their joint
distribution and their cross-asset dependence structure are modeled
explicitly. The second-stage likelihood treats the first-stage standardized innovations as
given, so the reported second-stage standard errors are conditional on the
first-stage filtering step.

\section{The Dynamic Correlation Matrix Modeling Framework \label{sec:Dynamic-Correlations-and}}

Let $Z_{t}=\left(Z_{t}^{N\prime},Z_{t}^{D\prime}\right)^{\prime}$
denote the $2n\times1$ vector of standardized innovations for all
$n$ assets, where $Z_{t}^{N}$ and $Z_{t}^{D}$ are the $n\times1$
vectors of overnight and intraday innovations, respectively. Building
on the asset-specific filtrations in Section \ref{sec:The-univariate-extended},
define
$
\mathcal{F}_{t}=\sigma (\left\{ R_{\tau}^{N},R_{\tau}^{D}\right\} _{\tau\leq t-1})
$ as the information available at the beginning
of day $t$, and 
$\mathcal{G}_{t}=\sigma(\mathcal{F}_{t}\cup\left\{ R_{t}^{N}\right\} )$ as the augmented filtration that incorporates the realized overnight returns prior to the intraday session.

In the most general setting, the joint conditional correlation matrix
of the $2n\times1$ innovation vector $Z_{t}$ can be partitioned
into a full block structure:
\[
C_{t}=\begin{pmatrix}C_{t}^{N} & C_{t}^{ND}\\
C_{t}^{DN} & C_{t}^{D}
\end{pmatrix}\in\mathbb{R}^{2n\times2n},
\]
where $C_{t}^{N}$ and $C_{t}^{D}$ are the overnight and
intraday correlation matrices, and $C_{t}^{ND}=(C_{t}^{DN})^{\prime}$
captures cross-session dependence. Details of the likelihood construction
and related derivations under this general specification are provided
in Appendix~\ref{sec:The-Score-and}. 

The first-stage Coupled EGARCH model removes the most direct same-asset
overnight-intraday dependence through the contemporaneous transmission
parameter $\delta_i$ in the mean equation and the cross-session volatility
spillovers. We then impose $C_t^{ND}=\mathbf{0}$ as a parsimonious block-diagonal
specification for the standardized innovations. This restriction is broadly consistent with the residual cross-session
correlation diagnostics reported in Appendix~\ref{sec:Diagnostic-Test-on},
where the remaining within-asset and cross-asset cross-session correlations
are small in magnitude and far below the within-session correlations.

Because the trading day unfolds sequentially, the augmented filtration
$\mathcal{G}_{t}$ is useful for describing the information available
before the intraday session. Under joint specifications such as the
common multivariate-$t$ distribution, the conditional distribution
of $Z_{t}^{D}$ may also depend on the realized overnight shocks $Z_{t}^{N}$.
In the block-diagonal session-specific specifications developed below,
this dependence is restricted through the assumed separation between
overnight and intraday innovation blocks. Accordingly, the conditional correlation matrix simplifies to $C_t=\operatorname{blockdiag}(C_t^N,C_t^D)\in\mathbb{R}^{2n\times 2n}$.

\begin{Definition}[Block Correlation Matrix]\label{def:BlockCorr}
    $C\in\mathbb{R}^{n\times n}$ is a block matrix
with $K$ blocks, if it is expressed as
\[
C=\left[\begin{array}{cccc}
C_{[1,1]} & C_{[1,2]} & \cdots & C_{[1,K]}\\
C_{[2,1]} & C_{[2,2]}\\
\vdots &  & \ddots\\
C_{[K,1]} &  &  & C_{[K,K]}
\end{array}\right], \quad \text{where } C_{[k,k]}=\left[\begin{array}{cccc}
1 & \rho_{kk} & \cdots & \rho_{kk}\\
\rho_{kk} & 1 & \ddots\\
\vdots & \ddots & \ddots\\
\rho_{kk} &  &  & 1
\end{array}\right],\ \ C_{[k,l]}=\left[\begin{array}{ccc}
\rho_{kl} & \cdots & \rho_{kl}\\
\vdots & \ddots\\
\rho_{kl} &  & \rho_{kl}
\end{array}\right]
\]
for $k\neq l$, and $\sum_{k=1}^{K}n_{k}=n$ with $n_k \geq 1$.
\end{Definition}

\subsection{The Unrestricted Parametrization of (Block) Correlation Matrix}
We model the overnight and intraday correlation matrices
using the score-driven framework by \citet{Creal2013}. Following
\citet{Archakov2021}, we parameterize each block of the conditional
correlation matrix via the matrix logarithm transformation. Let $\gamma_t^N=\operatorname{vecl}(\log C_t^N)\in\mathbb{R}^d$ and
$\gamma_t^D=\operatorname{vecl}(\log C_t^D)\in\mathbb{R}^d$, where
$d=n(n-1)/2$ and $\operatorname{vecl}(\cdot)$ selects and vectorizes the strictly lower triangular elements of the matrix. This transformation provides an unconstrained parameterization
of the correlation matrices and guarantees $C_{t}$ remains a unique
positive definite correlation matrix. Importantly, the matrix logarithm preserves block structures, as illustrated below
$$
\gamma(C)\equiv{\rm vecl}\left[\log\left(\begin{array}{cccc}
\cellcolor{black!5}1.0 & \cellcolor{black!5}0.6 & 0.2 & 0.2\\
\cellcolor{black!5}0.6 & \cellcolor{black!5}1.0 & 0.2 & 0.2\\
0.2 & 0.2 & \cellcolor{black!5}1.0 & \cellcolor{black!5}0.4\\
0.2 & 0.2 & \cellcolor{black!5}0.4 & \cellcolor{black!5}1.0
\end{array}\right)\right]={\rm vecl}\left[\begin{array}{cccc}
\cellcolor{black!5}-0.24 & \cellcolor{black!5}0.676 & 0.137 & 0.137\\
\cellcolor{black!5}0.676 & \cellcolor{black!5}-0.24 & 0.137 & 0.137\\
0.137 & 0.137 & \cellcolor{black!5}-0.11 & \cellcolor{black!5}0.404\\
0.137 & 0.137 & \cellcolor{black!5}0.404 & \cellcolor{black!5}-0.11
\end{array}\right]=\left(\begin{array}{c}
0.676\\
0.137 \times \iota_{4}\\
0.404
\end{array}\right),
$$
where $\iota_n$ is an $n \times 1$ vector of ones, and $C$ can be
$C_t^N$ or $C_t^D$.

For a block matrix with $K$ blocks as defined in Definition~\ref{def:BlockCorr}, $\gamma(C)$ will have at most $(K+1) K / 2$ distinct elements, such that we can write $\gamma=B \eta$, where $B$ is a known zero-one selector matrix and $\eta$ is a subvector of $\gamma$. In the example above we have, $\gamma={B} \eta$, ${B}= \operatorname{blockdiag}\left(1, \iota_4, 1\right) \in \mathbb{R}^{6 \times 3}$, and $\eta=(0.676,0.137,0.404)^{\prime}$. Parameterizing the block correlation matrix, $C$, with $\eta$ does not impose additional superfluous restrictions, see \cite{Tong2023}. Thus, any nonsingular block correlation matrix corresponds to a unique $\eta$ vector, and any dynamic block correlation model can be expressed as a dynamic model for $\eta$.

\subsection{The Score-driven Framework for Dynamic (Block) Correlation Matrix}

In the general case without imposing block structure, we can stack the parameters into a joint vector $\gamma_{t}=\left(\gamma_{t}^{N\prime},\gamma_{t}^{D\prime}\right)^{\prime}\in\mathbb{R}^{n(n-1)}$, and then model the dynamics of $\gamma_{t}$ via the score-driven framework of \cite{Creal2013} with a vector autoregressive model of order one, VAR(1):
\begin{equation}
\gamma_{t+1}=\left(I_{n(n-1)}-\beta\right)\mu+\beta\gamma_{t}+\alpha S_{t}^{-1}\nabla_{t},
\end{equation}
where $\mu$ is the unconditional mean of $\gamma_{t}$,
and $\alpha$ and $\beta$ are $n(n-1) \times n(n-1)$ diagonal matrices. The
vector $\nabla_{t}$ is the score of the joint log-likelihood, $\ell_{t}$,
taken with respect to $\gamma_{t}$:
\begin{equation}
\nabla_{t}=\frac{\partial\ell_{t}}{\partial\gamma_{t}}=\left(\frac{\partial\ell_{t}}{\partial\gamma_{t}^{N\prime}},\frac{\partial\ell_{t}}{\partial\gamma_{t}^{D\prime}}\right)^{\prime}\equiv\left(\nabla_{t}^{N\prime},\nabla_{t}^{D\prime}\right)^{\prime}.
\end{equation}

The scaling matrix $S_{t}$ is set to the conditional Fisher information
matrix, $S_{t}=\mathbb{E}_{t-1}\left[\nabla_{t}\nabla_{t}^{\prime}\right]$,
following \citet{Creal2013}. This choice makes the scaled score $S_{t}^{-1}\nabla_{t}$
an approximate Newton step in the parameter space, ensuring that the
magnitude of correlation updates is automatically adjusted for the
local curvature of the likelihood surface. To reduce computational
cost, we use a diagonal approximation of $S_{t}^{-1}$, whereby each
parameter is scaled by its own marginal Fisher information.

When a sector-based block structure is imposed on \(C_t^N\) and \(C_t^D\),
the dimension of \(\gamma_t^N\) and \(\gamma_t^D\) is reduced from
\(n(n-1)/2\) to \(K(K+1)/2\). From $\gamma_t^N=B^N\eta_t^N$ and $\gamma_t^D=B^D\eta_t^D$, we can stack them into a joint condensed vector $\eta_{t}=\left(\eta_{t}^{N\prime},\eta_{t}^{D\prime}\right)^{\prime}\in\mathbb{R}^{K\left(K+1\right)}$,
the score-driven dynamics for the condensed parameters become:
\begin{equation}
\eta_{t+1}=\left(I_{K(K+1)}-\beta\right)\mu+\beta\eta_{t}+\alpha S_{\eta,t}^{-1}\nabla_{\eta,t},
\end{equation}
Combined with the canonical representation in Subsection~\ref{sec:Block-structure},
the block structure allows the score and information matrices to be evaluated
in a lower-dimensional system governed by the number of blocks $K$ rather than
the number of assets $n$, which is desirable in high-dimensional
settings.  

The form of the score and Fisher information matrix depends on the distributional specifications introduced below.

\section{Multivariate Convolution-$t$ Distributions for Clustered Tail Heterogeneity
\label{sec:Convolution-t-Distributions}}

This section introduces the distributional component of the model. The convolution-$t$ framework of \citet{Hansen2026} provides a tractable class of
multivariate heavy-tailed distributions in which different subsets of variables
may have different degrees of freedom. It nests the multivariate-$t$ as a
special case. Flexible non-Gaussian specifications are particularly relevant in financial
covariance modeling, where departures from elliptical return distributions can
affect dependence modeling and portfolio decisions; see, for example,
\citet{PaolellaPolakWalker2021}. Our focus differs by using clustered
convolution-$t$ distributions to model session- and group-specific tail
heterogeneity in dynamic correlation systems.

A convolution-$t$ random vector is constructed by applying a correlation
transformation to several mutually independent multivariate $t$-distributed
components. Let $C_t^{1/2}$ denote the symmetric positive definite square root of $C_t$, and write 
$$
Z_{t}=C_{t}^{1/2}U_{t},\quad \text{where}\quad U_t=(U_{1,t}',\ldots,U_{G,t}')',\quad U_{g,t} \stackrel{\text { ind }}{\sim} t_{\nu_g}^{\mathrm{std}}(0,I_{m_g}),
$$
where $\nu_g>2$ and $\sum_{g=1}^G m_g=2n$.
Each cluster carries its own degrees-of-freedom parameter $\nu_{g}$,
so that tail thickness can vary across groups of variables while the
likelihood remains analytically tractable. 

The cluster partition can be chosen to reflect economic structure.
At one extreme, all variables share a single cluster ($G=1$), which
collapses to the standard multivariate-$t$. At the other extreme,
each asset in each session forms its own cluster ($G=2n,m_{g}=1$),
giving every variable its own tail parameter. In our setting, the
most natural partition groups variables by trading session, yielding
$G=2$ clusters, which means one for overnight innovations and one
for intraday innovations, each with its own degrees-of-freedom, $\nu_{N}$
and $\nu_{D}$. 

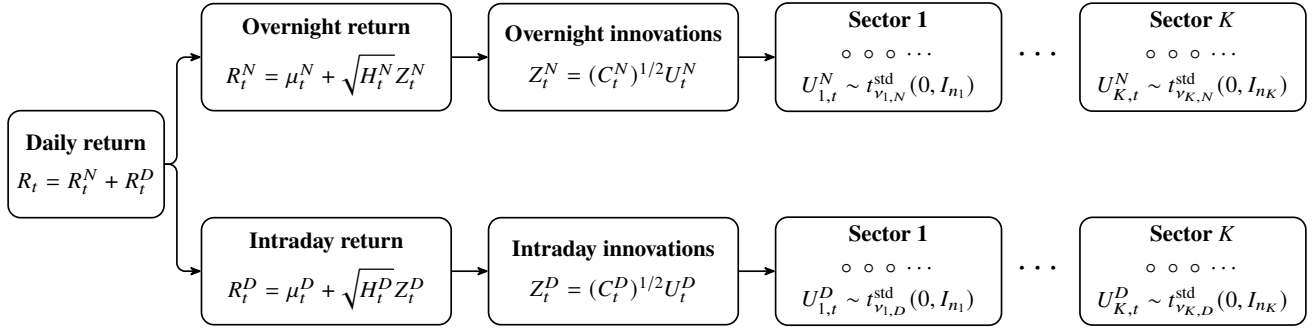
\begin{figure}
\begin{centering}
\tikzset{
	every node/.style={align=center, font=\small},
	card/.style={thick, rounded corners=6pt},
	daily/.style={draw=black, fill=white, card, minimum width=2.6cm, minimum height=1.8cm},
	step/.style={draw=black, fill=white, card, minimum width=4.2cm, minimum height=1.8cm},		
	sector/.style={draw=black, fill=white, card, minimum width=3.8cm, minimum height=1.8cm},
	dots/.style={draw=none, fill=none, font=\Large, minimum width=0.5cm},
	arr/.style={->, >={Stealth[round]}, thick, draw=black, rounded corners=4pt}
}

\resizebox{\textwidth}{!}{%
	\begin{tikzpicture}[node distance=0.6cm and 0.6cm]
		\node[daily] (daily) {
			\textbf{Daily return} \\[1ex] 
			$R_t=R_t^N+R_t^D$
		};
		
		\node[step=night, right=0.6cm of daily, yshift=1.8cm] (rn) {
			\textbf{Overnight return} \\[1.2ex]
			$R_t^N =\mu^{N}_t +\sqrt{H_t^N} Z_t^N$
		};
		
		\node[step=night, right=0.6cm of rn] (zn) {
			\textbf{Overnight innovations} \\[1.2ex]
			$Z_t^N = (C_t^N)^{1/2} U_t^N$
		};

		\node[sector=night, right=0.6cm of zn] (ns1) {
			\textbf{Sector 1} \\[0.4ex] 
			$\circ\ \circ\ \circ\ \cdots$ \\[0.4ex] 
			$U_{1,t}^{N}\sim t_{\nu_{1,N}}^{\mathrm{std}}(0,I_{n_{1}})$
		};
		
		\node[dots, right=0.10cm of ns1] (nsdots) {$\cdots$};
		
		\node[sector=night, right=0.10cm of nsdots] (nsk) {
			\textbf{Sector $K$} \\[0.4ex] 
			$\circ\ \circ\ \circ\ \cdots$ \\[0.4ex] 
			$U_{K,t}^{N}\sim t_{\nu_{K,N}}^{\mathrm{std}}(0,I_{n_{K}})$
		};

		\node[step=day, right=0.6cm of daily, yshift=-1.8cm] (rd) {
			\textbf{Intraday return} \\[1.2ex]
			$R_t^D = \mu^{D}_t+\sqrt{H_t^D} Z_t^D$
		};
		
		\node[step=day, right=0.6cm of rd] (zd) {
			\textbf{Intraday innovations} \\[1.2ex]
			$Z_t^D = (C_t^D)^{1/2} U_t^D$
		};

		\node[sector=day, right=0.6cm of zd] (ds1) {
			\textbf{Sector 1} \\[0.4ex] 
			$\circ\ \circ\ \circ\ \cdots$ \\[0.4ex] 
			$U_{1,t}^{D}\sim t_{\nu_{1,D}}^{\mathrm{std}}(0,I_{n_{1}})$
		};
		
		\node[dots, right=0.10cm of ds1] (dsdots) {$\cdots$};
		
		\node[sector=day, right=0.10cm of dsdots] (dsk) {
			\textbf{Sector $K$} \\[0.4ex] 
			$\circ\ \circ\ \circ\ \cdots$ \\[0.4ex] 
			$U_{K,t}^{D}\sim t_{\nu_{K,D}}^{\mathrm{std}}(0,I_{n_{K}})$
		};

		\draw[arr] (daily.east) -- ++(0.2,0) |- (rn.west);
		\draw[arr] (daily.east) -- ++(0.2,0) |- (rd.west);
		\draw[arr] (rn.east) -- (zn.west);
		\draw[arr] (zn.east) -- (ns1.west);
		\draw[arr] (rd.east) -- (zd.west);
		\draw[arr] (zd.east) -- (ds1.west);
	\end{tikzpicture}%
}
\par\end{centering}
\caption{Return Decomposition and Sessions/Sectoral Clustering \label{fig:Return-Decomposition-and-1}}
\end{figure}

More generally, the convolution-$t$ framework allows further flexible
partitioning. Assets within the overnight and intraday sessions could
be grouped into multiple clusters, each associated with its own tail
parameter. For instance, assets could be grouped by economic industry
sectors, allowing tail behavior to vary across both sessions and economic
sectors simultaneously.  Figure \ref{fig:Return-Decomposition-and-1} illustrates this structure:
daily returns are first decomposed into overnight and intraday components,
and each session is then partitioned into $K$ sector-based clusters,
with each cluster carrying its own degrees-of-freedom parameter $\nu_{k,c}$,
for $k=1, \ldots,K$ and $c\in\left\{ N,D\right\} $. The total number
of clusters is thus $G=2K$. We consider all of these specifications
in the empirical analysis.

Since the clusters $U_{g,t}$ are mutually independent, the joint
log-likelihood is additively separable:
\begin{equation}
\ell_{t}\left(Z_{t}\mid\mathcal{F}_{t}\right)=-\frac{1}{2}\log|C_{t}|+\sum_{g=1}^{G}\ell_{g,t}\left(U_{g,t};\nu_{g}\right),
\end{equation}

\noindent where $\ell_{g,t}$ denotes the log-density contribution
of the $g$-th cluster.

\subsection{Case 1: Multivariate $t$-Distribution (MT)}

Under the standardized multivariate-$t$ distribution, $Z_t\mid\mathcal{F}_t
\sim t^{\mathrm{std}}_{2n,\nu}(0,C_t)$, with log-likelihood given by
\begin{equation}
    \ell_{t}\left(Z_{t}\mid\mathcal{F}_{t}\right)=c_{\nu,2n}-\frac{1}{2}\log|C_{t}|-\frac{\nu+2n}{2}\log\left(1+\frac{1}{\nu-2}Z_{t}^{\prime}C_{t}^{-1}Z_{t}\right),
\end{equation}
where $c_{\nu,2n}$ is the normalizing constant, and the
scaling by $\nu-2$ ensures that the distribution has unit variance
for $\nu>2$. The Gaussian specification is nested as the limiting case $\nu\to\infty$,
under which the tail adjustment disappears and the multivariate-$t$
likelihood reduces to the Gaussian likelihood.

Because all $2n$ elements share a common degrees-of-freedom parameter
$\nu$, the multivariate-$t$ specification imposes a homogeneous tail
structure across overnight and intraday innovations.

By the marginal-conditional representation of the standardized multivariate-$t$
distribution,
\[
Z_t^N\mid \mathcal{F}_t
\sim t_\nu^{\mathrm{std}}(0,C_t^N),\qquad
Z_t^D\mid \mathcal{G}_t
\stackrel{d}{=}
Z_t^D\mid Z_t^N
\sim t_{\nu+n}^{\mathrm{std}}
\left(C_t^{DN}(C_t^N)^{-1}Z_t^N,\widetilde C_t^D\right),
\]
where
\[
\widetilde C_t^D
=
\frac{\nu-2+q_N}{\nu+n-2}
\left(
C_t^D-C_t^{DN}(C_t^N)^{-1}C_t^{ND}
\right),
\qquad
q_N=(Z_t^N)'(C_t^N)^{-1}Z_t^N .
\]
This implies the decomposition $\ell_{t}=\ell\left(Z_{t}^{N}\mid\mathcal{F}_{t}\right)+\ell\left(Z_{t}^{D}\mid\mathcal{G}_{t}\right)$. However, the two components remain jointly governed by the same degrees-of-freedom
parameter $\nu$, and the conditional distribution of $Z_{t}^{D}$
depends on $Z_{t}^{N}$. Thus, the model imposes a homogeneous tail
structure across sessions.

Under the block-diagonal specification $C_t^{ND}=\mathbf{0}$, this reduces to
\[
Z_t^D\mid\mathcal{G}_t
\sim t_{\nu+n}^{\mathrm{std}}
\left(
0,
\frac{\nu-2+q_N}{\nu+n-2}C_t^D
\right),
\qquad
q_N=(Z_t^N)'(C_t^N)^{-1}Z_t^N .
\]

Thus, even when $C_t^{ND}=\mathbf{0}$, the conditional intraday distribution is scaled
by the realized overnight shock and both sessions remain governed by the same
tail parameter. This motivates the convolution-$t$ specifications below, which
allow tail behavior to differ across sessions while preserving tractability.

\subsection{Case 2: Cluster-$t$ Distributions}

The Cluster-$t$ specification arises when $G$ is determined by a
predefined grouping structure. In this case, each subvector $U_{g,t}$
follows a multivariate $t$-distribution, and variables within the
same cluster share a common degrees-of-freedom parameter $\nu_{g}$.
Aggregating the cluster-specific
contributions gives
\begin{equation}
\ell_t\left(Z_t\mid\mathcal{F}_t\right)
=
-\frac{1}{2}\log|C_t|
+
\sum_{g=1}^{G}
\left[
c_{\nu_g,m_g}
-\frac{\nu_g+m_g}{2}
\log\left(
1+\frac{U_{g,t}^{\prime}U_{g,t}}{\nu_g-2}
\right)
\right],
\end{equation}
where $c_{\nu_g,m_g}$ is the normalizing constant for the $g$-th cluster.

One important feature of the Cluster-$t$ specification is that it
preserves the nonlinear dependence among variables within each cluster.
Because variables in the same cluster share a common tail thickness,
the model can capture common extreme realizations within a sector
or within a trading session.

\subsection{Case 3: Hetero-$t$ Distributions}

The Hetero-$t$ is the special case in which each element of the innovation
vector forms its own cluster, so that $G=2n$ and $m_{g}=1$. This
specification allows each variable to have its own degrees-of-freedom,
providing the most flexible specification for heterogeneous tail behavior.
Under this limiting case, every asset $j$ in each session $c$ carries
a unique parameter $\nu_{j,c}$, resulting in $G=2n$ independent
univariate clusters. Thus, the log-likelihood is given by 
\begin{equation}
\ell_{t}\left(Z_{t}\mid\mathcal{F}_{t}\right)=-\frac{1}{2}\log|C_{t}|+\sum_{g=1}^{2n}\left[c_{\nu_{g},1}-\frac{\nu_{g}+1}{2}\log\left(1+\frac{1}{\nu_{g}-2}U_{g,t}^{2}\right)\right],
\end{equation}

\noindent where $c_{\nu_{g},1}$ is the normalizing constant, and
$U_{g,t}$ is the $g$-th element of the standardized vector $U_{t}$. However, this flexibility comes at a cost. When $m_{g}=1$, the components
of $U_{t}$ are mutually independent, implying that no joint tail
dependence is preserved prior to the correlation transformation. As
a result, the Hetero-$t$ specification captures marginal heavy tails
but does not preserve a shared cluster-level tail component before
the correlation transformation. It therefore provides a less direct
representation of clustered extreme co-movements than the Cluster-$t$
specification.

The choice between Cluster-$t$ and Hetero-$t$ therefore reflects
a trade-off: the former preserves within-group tail dependence while
allowing for moderate heterogeneity, whereas the latter maximizes
marginal flexibility at the expense of weakening the cluster-level
tail structure. Thus, the convolution-$t$ framework provides a continuum of heavy-tailed
multivariate specifications, ranging from a common-tail elliptical model to
fully heterogeneous marginal tails, with intermediate cluster structures that
preserve within-group tail dependence.

\section{The Scores and Fisher Information under Convolution-$t$ Distributions \label{subsec:Scores-and-Fisher}}

In this section, we derive the score $\nabla_{t}$ and the Fisher
information matrix $\mathcal{I}_{t}$ under each of the distributional
specifications. The specifications differ primarily in how extreme
observations are weighted when updating the correlation parameters,
with the multivariate-$t$ using a single global weight and the Cluster-$t$
and Hetero-$t$ using cluster- and asset-specific weights, respectively.
The derivations below build on \citet{Tong2026}, extending their
framework to accommodate the block structure of the joint innovation
vector.

To facilitate the derivations, we define the Jacobian matrix of the
logarithmic transformation as $M_{t}=\partial\operatorname{vec}\left(C_{t}\right)/\partial\gamma_{t}^{\prime}$.
Given the block-diagonal structure of the joint correlation matrix,
$C_{t}=\operatorname{blockdiag}\left(C_{t}^{N},C_{t}^{D}\right)$, we introduce
a $4n^{2}\times2n^{2}$ embedding matrix $P_{M}$ that inserts rows of zeros corresponding to the cross-block zero entries. The full
Jacobian is then constructed as $M_{t}=P_{M}\operatorname{blockdiag}\left(M_{t}^{N},M_{t}^{D}\right)$,
where $M_{t}^{N}=\partial\operatorname{vec}\left(C_{t}^{N}\right)/\partial\gamma_{t}^{N\prime}$
and $M_{t}^{D}=\partial\operatorname{vec}\left(C_{t}^{D}\right)/\partial\gamma_{t}^{D\prime}$
are the Jacobians of the respective $n\times n$ sub-blocks. $\otimes$ is the Kronecker product. We use the shorthand \(A\oplus B=A\otimes B+B\otimes A\), as in
\citet{Creal2011}.

\subsection{Case 1: The Multivariate-$t$ Specification\label{subsec:The-Multivariate--benchmark}}

If $Z_{t}$ follows a $2n$-dimensional (standardized) multivariate-$t$
distribution with a single degree-of-freedom $\nu$, the score with
respect to $\gamma_{t}$ takes the form:
\begin{equation}
\nabla_{t}^{MT}=\frac{1}{2}M_{t}^{\prime}\left(C_{t}^{-1}\otimes C_{t}^{-1}\right)\left[W_{t}\operatorname{vec}\left(Z_{t}Z_{t}^{\prime}\right)-\operatorname{vec}\left(C_{t}\right)\right],\label{eq:score-MT}
\end{equation}
where $W_t=(\nu+2n)/(\nu-2+Z_t^\prime C_t^{-1}Z_t)$ acts as a global
scaling factor that down-weights large Mahalanobis distances in the score. The Gaussian specification
arises as the limiting case $\nu\to\infty$ under which $W_{t}\to1$. Unlike the multivariate-$t$ case, the Gaussian score assigns uniform
weight to all observations regardless of the realized Mahalanobis
distance, making it unbounded and sensitive to extreme shocks.

Due to the block-diagonal structure of $C_{t}$, the joint score under
both specifications partitions into overnight and intraday components:
\begin{equation}
\nabla_{t}=\left(\nabla_{t}^{N\prime},\nabla_{t}^{D\prime}\right)^{\prime}, \quad \text{where}\quad \nabla_{t}^{c}=\tfrac{1}{2}\left(M_{t}^{c}\right)^{\prime}\left(\left(C_{t}^{c}\right)^{-1}\otimes\left(C_{t}^{c}\right)^{-1}\right)\left[W_{t}\operatorname{vec}\left(Z_{t}^{c}\left(Z_{t}^{c}\right)^{\prime}\right)-\operatorname{vec}\left(C_{t}^{c}\right)\right],
\end{equation}
for $c\in\{N,D\}$. Under the multivariate-$t$ distribution, both $\nabla_{t}^{N}$
and $\nabla_{t}^{D}$ share the scaling weight $W_{t}$, which incorporates
both $Z_{t}^{N}$ and $Z_{t}^{D}$. An extreme overnight shock would
decrease the total weight $W_{t}$. This cross-session coupling may
lead to inefficient score updates when tail risks differ substantially
across trading sessions.

The total Fisher information matrix decomposes as:
\begin{equation}
\mathcal{I}_{t}^{MT}=\left(\begin{array}{cc}
\mathcal{I}_{N} & \mathbf{0}\\
\mathbf{0} & \mathcal{I}_{D}
\end{array}\right)+\tfrac{\phi-1}{4}\mathcal{I}_{ND}\mathcal{I}_{ND}^{\prime},\quad \text{where}\quad \mathcal{I}_{c}=\frac{\phi}{4}\left(M_{t}^{c}\right)^{\prime}\left[\left(\left(C_{t}^{c}\right)^{-1}\otimes\left(C_{t}^{c}\right)^{-1}\right)H_{n}\right]M_{t}^{c},
\end{equation}
for $c\in\{N,D\}$, with $\phi=\frac{\nu+2n}{\nu+2n+2}$ under the multivariate-$t$
distribution and $\phi=1$ under the Gaussian. $H_{n}=I_{n^{2}}+K_{n}$ and $K_{n}$ is the commutation matrix. The cross-session term is given by $\mathcal{I}_{ND}=M_{t}^{\prime}\operatorname{vec}\left(C_{t}^{-1}\right)$.
Under the Gaussian specification, $\phi=1$ and the coupling term
$\frac{\phi-1}{4}\mathcal{I}_{ND}\mathcal{I}_{ND}^{\prime}$ vanishes
identically, so the Fisher information is exactly block-diagonal and
the two sessions are fully decoupled.

\subsection{Case 2: The Cluster-$t$ Specification\label{subsec:The-Cluster--Specification}}

The Cluster-$t$ specification addresses the coupling problem in multivariate-$t$
by allowing a predefined partition of assets into independent clusters,
each carrying its own degrees-of-freedom parameter. In its simplest
form, this specification groups variables by trading session, treating
overnight and intraday innovations as two independent multivariate-$t$
vectors. Under this session-level partition, the score for each session
$c\in\{N,D\}$ follows the same functional form as the multivariate-$t$
in Section \ref{subsec:The-Multivariate--benchmark}, but is updated
using a session-specific scaling weight $W_{t}^{c}$ and degrees-of-freedom
$\nu_{c}$. This ensures that an extreme overnight shock no longer
distorts the intraday parameter updates.

More generally, our framework allows for a more precise partition
within each session. Suppose the $n$ assets within session $c\in\{N,D\}$
are grouped into $K$ sector-based clusters, with sector $k$
containing $n_{k}$ assets and $\sum_{k=1}^{K}n_{k}=n$. Define the orthogonalized shocks as $U_{t}^{c}=(C_{t}^{c})^{-1/2}Z_{t}^{c}$ and partition them into
$K$ subvectors $U_{k,t}^{c}$. Each subvector follows an independent
multivariate-$t$ distribution with session- and sector-specific degrees-of-freedom
parameter $\nu_{k,c}$.

Following Theorem 2 of \citet{Tong2026}, the score for session $c$
takes the form:
\begin{equation}
\nabla_{t}^{c,CT}
=
\left(M_{t}^{c}\right)^{\prime}
\left(\Omega_{t}^{c}\right)^{\prime}
\left[
\sum_{k=1}^{K}
W_{k,t}^{c}\,
\operatorname{vec}\left(E_{k}U_{k,t}^{c}\left(U_{t}^{c}\right)^{\prime}\right)
-\operatorname{vec}\left(I_{n}\right)
\right],\quad \text{where}\quad W_{k,t}^{c}=\frac{\nu_{k,c}+n_{k}}{\nu_{k,c}-2+(U_{k,t}^{c})^{\prime}U_{k,t}^{c}}
\end{equation}
for $k=1, \ldots,K$, \(E_k\in\mathbb{R}^{n\times n_k}\) is the embedding matrix that maps the \(k\)-th sector subvector into the full \(n\)-dimensional vector, so that \(E_k'\) selects the \(k\)-th sector from the full vector, and $\Omega_{t}^{c}=(I_{n}\otimes(C_{t}^{c})^{-1/2})((C_{t}^{c})^{1/2}\oplus I_{n})^{-1}$.

This localized weighting scheme is the key feature of the Cluster-$t$
specification. An extreme shock in sector $k$ reduces the sector-specific $W_{k,t}^{c}$ and hence down-weights the score contribution associated
with that sector, rather than globally down-weighting all sector contributions. Under
the multivariate-$t$, by contrast, a single extreme shock depresses
the global weight $W_{t}$ and distorts the score updates for all
assets simultaneously. When $K=1$, the score collapses to the session-level
case with a single weight $W_{t}^{c}=\frac{\nu_{c}+n}{\nu_{c}-2+\left(U_{t}^{c}\right)^{\prime}U_{t}^{c}}$
for $c\in\{N,D\}$.

Since the overnight and intraday cluster systems are independent under the session-separated convolution-$t$ specification, the Fisher information is block-diagonal across sessions:
\begin{equation}
\mathcal{I}_{t}^{CT}=\begin{pmatrix}\mathcal{I}_{N}^{CT} & \mathbf{0}\\
\mathbf{0} & \mathcal{I}_{D}^{CT}
\end{pmatrix},\quad \text{where}\quad \mathcal{I}_{c}^{CT}
=
\left(M_{t}^{c}\right)^{\prime}
\left(\Omega_{t}^{c}\right)^{\prime}
\left(K_{n}+\Upsilon_{K}^{c}\right)
\Omega_{t}^{c}M_{t}^{c}, \quad \Upsilon_{K}^{c}=\sum_{k=1}^{K}\Psi_{k}^{c}
\end{equation}
for $c\in\left\{ N,D\right\}$, where $\Psi_{k}^{c}=\psi_{k}^{c}\left(I_{n}\otimes J_{k}\right)+(\phi_{k}^{c}-\psi_{k}^{c})(J_{k}\otimes J_{k})+(\phi_{k}^{c}-1)[(J_{k}\otimes J_{k})K_{n}+\operatorname{vec}(J_{k})\operatorname{vec}(J_{k})^{\prime}]$, where $J_{k}=E_{k}E_{k}^{\prime}$, $\phi_{k}^{c}=\tfrac{\nu_{k,c}+n_{k}}{\nu_{k,c}+n_{k}+2}$,
and $\psi_{k}^{c}=\phi_{k}^{c}\tfrac{\nu_{k,c}}{\nu_{k,c}-2}$. This block-diagonal structure follows directly from the mutual
independence of clusters across sessions imposed in Section
\ref{sec:Convolution-t-Distributions}.

\subsection{Case 3: The Hetero-$t$ Specification}

The Hetero-$t$ distribution is the limiting case in which each asset
within each session forms its own independent cluster ($K=n$, $n_{k}=1$
for all $k$). This gives every asset its own degrees-of-freedom parameter
$\nu_{j,c}$, providing maximum flexibility in tail modeling. 

The score takes the same form as in Section \ref{subsec:The-Cluster--Specification}
with $K=n$. Since each cluster contains a single asset, the sector-level
weight reduces to
$W_{j,t}^c=(\nu_{j,c}+1)/\{\nu_{j,c}-2+(U_{j,t}^c)^2\}$,
$j=1,\ldots,n$. Here, $U_{j,t}^{c}$ is the $j$-th element of $U_{t}^{c}$. The score then becomes:
\begin{equation}
\nabla_{t}^{c,HT}
=
\left(M_{t}^{c}\right)^{\prime}
\left(\Omega_{t}^{c}\right)^{\prime}
\left[
\sum_{j=1}^{n}
W_{j,t}^{c}\,
\operatorname{vec}\left(E_{j}U_{j,t}^{c}\left(U_{t}^{c}\right)^{\prime}\right)
-\operatorname{vec}\left(I_{n}\right)
\right],
\end{equation}

\noindent where $E_{j}$ is the $n\times1$ selection vector with
unity in the $j$-th position and zeros elsewhere. 

The Fisher information matrix retains the same block-diagonal structure
as in the Cluster-$t$ specification. The inner matrix $\Upsilon_{n}^{c}$
simply adapts to the asset-level partition as $\Upsilon_{n}^{c}=\sum_{j=1}^{n}\Psi_{j}^{c}$,
where $\Psi_{j}^{c}$ is computed using the formulas from Section
\ref{subsec:The-Cluster--Specification} by setting $n_{k}=1$ and
replacing the sector index $k$ with the asset index $j$.

\subsection{The Key to High Dimensions: Canonical Representation of Block Structure\label{sec:Block-structure}}

The analytical scores and information matrices derived above are formulated for the unrestricted
$n\times n$ session-specific correlation matrices, $C_{t}^{N}$ and
$C_{t}^{D}$. However, when the number of assets $n$ is large, updating
the unrestricted parameter vectors $\gamma_{t}^{N},\gamma_{t}^{D}\in\mathbb{R}^{n\left(n-1\right)/2}$
is still computationally difficult. Although the block structure reduces the number of free parameters
per session, computing the condensed score $\nabla_{\eta,t}^{c}$
still requires differentiating through the full $n\times n$ correlation
matrix. We employ the canonical representation derived in \citet{Archakov2024}, where any block correlation matrix can be decomposed
as $C_{t}^{c}=QD_{t}^{c}Q^{\prime}$. Here, $Q$ is a time-invariant
orthogonal matrix determined solely by the cluster sizes, and $D_{t}^{c}$
is a block-diagonal matrix taking the form:
\begin{equation}
D_{t}^{c}=\operatorname{blockdiag}\left(A_{t}^{c},\lambda_{1,t}^{c}I_{n_{1}-1},\dots,\lambda_{K,t}^{c}I_{n_{K}-1}\right).
\end{equation}

Since the eigenvalues $\lambda_{k,t}^{c}$ are deterministic functions
of the elements in $A_{t}^{c}$, the dynamic properties of the entire
$n\times n$ correlation matrix $C_{t}^{c}$ are entirely captured
by the lower-dimensional $K\times K$ symmetric positive definite
matrix $A_{t}^{c}$, for $c\in\{N,D\}$.  The chain rule gives
$\nabla_{\eta,t}^c=\Pi_{A,c}'\nabla_{A,t}^c$ for $c\in\{N,D\}$,
where $\nabla_{A,t}^c=\partial\ell_t^c/\partial\operatorname{vec}(A_t^c)$
is the score with respect to $A_t^c$. Closed-form expressions for
both quantities under the multivariate-$t$, Cluster-$t$, and Hetero-$t$
specifications are provided in Theorems 3-5 of \citet{Tong2026}.
The transition matrix $\Pi_{A,c}=\frac{\partial\operatorname{vec}\left(A_{t}^{c}\right)}{\partial\eta_{t}^{c\prime}}$
is given by:
\begin{equation}
\Pi_{A,c}=\left[\Gamma_{A,c}-\Gamma_{A,c}E_{d}^{\prime}\left(\Phi_{c}^{-1}+E_{d}\Gamma_{A,c}E_{d}^{\prime}\right)^{-1}E_{d}\Gamma_{A,c}\right]\left(\Lambda_{n}\otimes\Lambda_{n}\right)D_{K},
\end{equation}
where $\Gamma_{A,c}=\partial\operatorname{vec}(A_t^c)/
\partial\operatorname{vec}(\log A_t^c)$, $\Phi_c$ is a $K\times K$ diagonal matrix with $\Phi_{c,kk}=\lambda_{k,t}^{c}\left(n_{k}-1\right)$,
and $\Lambda_{n}=\operatorname{diag}(\sqrt{n_{1}}, \ldots,\sqrt{n_{K}})$,
$E_{d}$ is an elimination matrix, and $D_{K}$ is the duplication
matrix. Evaluating $\Pi_{A,c}$ requires inverting only a $K\times K$
matrix, which keeps the model computationally tractable even in high
dimensions.

Stacking the session-specific components gives
$\nabla_{\eta,t}=(\nabla_{\eta,t}^{N\prime},\nabla_{\eta,t}^{D\prime})^\prime$,
with $\nabla_{\eta,t}^c=\Pi_{A,c}'\nabla_{A,t}^c$ and
$\mathcal{I}_{\eta,t}^{c}=\Pi_{A,c}^{\prime}\mathcal{I}_{A,t}^{c}\Pi_{A,c}$, where $c\in\{N,D\}$.  For the session-separated convolution-\(t\) specifications, 
\(\mathcal I_{\eta,t}\) remains block-diagonal, so the overnight and intraday
parameters can be updated separately. In the common multivariate-\(t\) case,
the additional coupling term in the Fisher information remains.

\section{Benchmark Model: DCC-GARCH Model \label{sec:Benchmark-DCC}}

We use the Dynamic Conditional Correlation (DCC) GARCH model of \citet{Engle2002},
implemented through the corrected DCC (cDCC) specification of \citet{Aielli2013},
as a benchmark. We adapt the cDCC specification to the joint overnight-intraday
innovation vector $Z_t=(Z_t^{N\prime},Z_t^{D\prime})^{\prime}$ obtained from the
first-stage Coupled EGARCH model. In the cDCC framework, the conditional
correlation matrix is formulated as
\[
C_t=\Lambda_{Q_t}^{-1/2}Q_t\Lambda_{Q_t}^{-1/2},
\]
where $Q_t$ is a $2n\times 2n$ symmetric positive definite matrix, and
$\Lambda_{Q_t}$ is the diagonal matrix containing the diagonal elements of
$Q_t$. The dynamic properties of $C_t$ are governed by the evolution of $Q_t$,
which is updated via
\begin{equation}
Q_{t+1}
=
(\iota\iota^{\prime}-\alpha-\beta)\odot\bar C
+
\beta\odot Q_t
+
\alpha\odot
\left(
\Lambda_{Q_t}^{1/2}Z_tZ_t^{\prime}\Lambda_{Q_t}^{1/2}
\right),
\label{eq:cDCC}
\end{equation}
where $\iota$ is a $2n\times1$ vector of ones, $\odot$ denotes the Hadamard
product, and $\bar C$ is the unconditional sample correlation matrix of $Z_t$.
The parameter matrices $\alpha$ and $\beta$ capture the innovation impact and
persistence of the correlation dynamics, respectively. The diagonal elements of
$C_t$ are normalized to one by construction.

To make the benchmark comparable with the score-driven models, we evaluate the
DCC model under the same main distributional specifications. Under the Gaussian
benchmark, $Z_t\mid\mathcal F_t\sim N(0,C_t)$. Under the multivariate-$t$
benchmark, $Z_t\mid\mathcal F_t\sim t^{\mathrm{std}}_{2n,\nu}(0,C_t)$, so all
overnight and intraday innovations share a common degrees-of-freedom parameter.
Under the session-level Cluster-$t$ benchmark, we write $Z_t=C_t^{1/2}U_t$,
where $U_t=(U_t^{N\prime},U_t^{D\prime})^{\prime}$, $
U_t^N\sim t^{\mathrm{std}}_{n,\nu_N}(0,I_n)$, $U_t^D\sim t^{\mathrm{std}}_{n,\nu_D}(0,I_n)$, and the two session components are mutually independent. This specification
allows overnight and intraday innovations to have different tail thickness while
retaining the DCC correlation update.

\section{Six-Asset Empirical Application\label{sec:Six-Asset-Application}}

This section presents a detailed empirical application using six U.S. equities
from two industrial sectors. The small cross section allows us to examine the
main mechanisms of the proposed model in detail: session-level tail
heterogeneity, sector-level tail clustering, unrestricted versus block
correlation dynamics, and out-of-sample predictive performance.

\subsection{Data and Descriptive Statistics\label{subsec:Six-Asset-Data}}

We begin with six major U.S. stocks from two industrial sectors. The Energy
sector is represented by Chevron (CVX), APA Corporation (APA), and Devon Energy
(DVN), and the Information Technology sector by Microsoft (MSFT), Intel (INTC),
and Cisco (CSCO). The two sectors differ in their information environments:
energy stocks are more exposed to global commodity and macroeconomic news,
while technology stocks are more sensitive to firm-specific announcements,
including earnings and guidance releases that may occur outside regular trading
hours. This contrast motivates the sector-based block structure imposed on the
correlation matrix and the sector-based clustering in the convolution-$t$
distribution. The sample spans from January 1, 2005, to December 31, 2024, with
a total of $T=5{,}032$ observations. Daily stock price data are from the Center
for Research in Security Prices (CRSP) database, and all total, overnight, and
intraday returns are computed as logarithmic returns.

\begin{table}
\caption{Descriptive Statistics\label{tab:Descriptive-Statistics}}
\centering{}\begin{threeparttable}
\footnotesize
\begin{tabularx}{\textwidth}{l l YYYY}
\toprule
\midrule
Asset & Session & Mean (Ann., \%) & Std. Dev. (Ann., \%) & Skewness & Kurtosis \\
\midrule
CVX & Total & 5.2371 & 28.5322 & -0.5027 & 24.4274 \\
 & Night & 5.2930 & 15.9007 & -1.0612 & 34.0504 \\
 & Day & -0.0559 & 22.3116 & -0.3489 & 17.4820 \\
\cmidrule{1-6}
APA & Total & -3.6940 & 50.1603 & -3.1880 & 84.7365 \\
 & Night & 15.2066 & 28.1480 & -2.4121 & 92.1604 \\
 & Day & -18.9005 & 39.4028 & -0.8950 & 17.1931 \\
\cmidrule{1-6}
DVN & Total & -0.6411 & 45.3060 & -0.8529 & 22.1656 \\
 & Night & 13.7665 & 25.4334 & -3.0507 & 77.1893 \\
 & Day & -14.4076 & 36.3537 & -0.0588 & 7.3408 \\
\cmidrule{1-6}
MSFT & Total & 13.8102 & 27.1190 & -0.0553 & 12.4788 \\
 & Night & 6.3938 & 16.0672 & -0.3731 & 28.4753 \\
 & Day & 7.4164 & 21.1623 & 0.0025 & 7.5876 \\
\cmidrule{1-6}
INTC & Total & -0.7026 & 33.0562 & -1.0092 & 19.2634 \\
 & Night & -4.6812 & 20.2724 & -3.6139 & 69.5382 \\
 & Day & 3.9786 & 25.2623 & 0.1255 & 6.7911 \\
\cmidrule{1-6}
CSCO & Total & 5.6078 & 28.0333 & -0.4621 & 15.2679 \\
 & Night & -1.7776 & 18.2802 & -1.7500 & 43.2764 \\
 & Day & 7.3854 & 21.1650 & -0.1789 & 8.2608 \\
\bottomrule
\end{tabularx}
\begin{tablenotes}[flushleft]
\footnotesize
\item \textit{Note}: This table reports descriptive statistics. Mean and Std. Dev. are annualized percentages, obtained by multiplying the sample mean by $252\times100$ and the sample standard deviation by $\sqrt{252}\times100$, respectively. Kurtosis denotes raw kurtosis. Total, Overnight, and Intraday refer to close-to-close, close-to-open, and open-to-close returns.
\end{tablenotes}
\end{threeparttable}
\end{table}

Table \ref{tab:Descriptive-Statistics} reports the descriptive statistics.
Average returns are small at the daily frequency, but annualized means reveal
clear session-level and sectoral differences. Energy stocks have higher average
overnight returns than intraday returns: APA shows an annualized overnight mean
of 15.21\% versus an intraday mean of -18.90\%, and DVN shows 13.77\% versus
-14.41\%. Technology stocks display more heterogeneous patterns, with MSFT
earning positive average returns in both sessions and CSCO showing a negative
overnight mean but a positive intraday mean.

Intraday returns have higher annualized standard deviations than overnight
returns for all six assets. For example, CVX has an intraday standard deviation
of 22.31\% compared with 15.90\% overnight, and MSFT has 21.16\% intraday
versus 16.07\% overnight. This reflects the concentration of trading activity
during market hours.

The most pronounced difference across sessions appears in tail behavior.
Overnight kurtosis is extreme, reaching 92.16 for APA, 77.19 for DVN, and
69.54 for INTC, whereas intraday kurtosis is much lower, ranging from 6.79
(INTC) to 17.49 (CVX). This session-level difference in tail thickness is
consistent with the tail heterogeneity that motivates the convolution-$t$
framework. Overnight returns also tend to be more negatively skewed, but the
main empirical feature for our model is the stronger heavy-tailed behavior of
overnight returns.

Table \ref{tab:Session-Correlation} presents the static correlation matrices of
the standardized innovations, with overnight innovation correlations in the
upper triangle and intraday innovation correlations in the lower triangle.

\begin{table}
\caption{Session Correlations of Standardized Innovations\label{tab:Session-Correlation}}
\centering{}\begin{threeparttable}
	\footnotesize
	\setlength{\tabcolsep}{3pt}
	
	\begin{tabularx}{\textwidth}{l YYYYYY}
		\toprule
		\midrule
		& \multicolumn{3}{c}{\textbf{Energy Sector}} & \multicolumn{3}{c}{\textbf{Technology Sector}} \\
		\cmidrule(lr){2-4} \cmidrule(lr){5-7}
		& CVX & APA & DVN & MSFT & INTC & CSCO \\
		\midrule
		CVX  & 1.000 & 0.780 & 0.809 & \cellcolor{black!8}0.430 & \cellcolor{black!8}0.407 & \cellcolor{black!8}0.450 \\
		APA  & 0.604 & 1.000 & 0.834 & \cellcolor{black!8}0.378 & \cellcolor{black!8}0.352 & \cellcolor{black!8}0.379 \\
		DVN  & 0.656 & 0.729 & 1.000 & \cellcolor{black!8}0.407 & \cellcolor{black!8}0.370 & \cellcolor{black!8}0.403 \\
		\addlinespace[0.5ex]
		MSFT & \cellcolor{black!8}0.382 & \cellcolor{black!8}0.252 & \cellcolor{black!8}0.272 & 1.000 & 0.483 & 0.454 \\
		INTC & \cellcolor{black!8}0.363 & \cellcolor{black!8}0.261 & \cellcolor{black!8}0.289 & 0.554 & 1.000 & 0.416 \\
		CSCO & \cellcolor{black!8}0.409 & \cellcolor{black!8}0.295 & \cellcolor{black!8}0.324 & 0.582 & 0.555 & 1.000 \\
\midrule
\bottomrule
	\end{tabularx}
	
	\begin{tablenotes}[flushleft]
		\footnotesize
		\item \textit{Note}: This table presents the sample correlation matrices of
standardized innovations for six assets. The upper triangular part displays
correlations of standardized overnight innovations; the lower triangular part
displays correlations of standardized intraday innovations. Off-diagonal sector
blocks correspond to cross-sector correlations.
	\end{tablenotes}
\end{threeparttable}
\end{table}

Within-sector correlations are substantially higher than cross-sector
correlations in both sessions. For the Energy sector, within-sector
correlations range from 0.780 to 0.834 for overnight innovations and from 0.604
to 0.729 for intraday innovations. For the Information Technology sector,
within-sector correlations range from 0.416 to 0.483 for overnight innovations
and from 0.554 to 0.582 for intraday innovations. Cross-sector correlations are
lower in both sessions, ranging from 0.352 to 0.450 overnight and from 0.252 to
0.409 intraday.

The session-level pattern differs between sectors. For Energy, overnight
innovation correlations exceed intraday innovation correlations across all
within-sector pairs. For Technology, the pattern reverses: intraday innovation
correlations exceed overnight innovation correlations. Cross-sector correlations
are moderately higher overnight than intraday. These static estimates provide a preliminary characterization of the remaining dependence among standardized innovations. The model-based dynamic conditional correlations are discussed in
Section \ref{subsec:Six-Asset-Multivariate}.

\subsection{Univariate Model Estimates\label{subsec:Six-Asset-Univariate}}

We estimate the Coupled EGARCH model asset by asset via QMLE. 
Table~\ref{tab:Mean-Equation} reports the conditional mean estimates with
Parzen-kernel HAC standard errors, while Table~\ref{tab:Volatility-Dynamics}
reports the volatility-parameter estimates with robust QMLE standard errors.
Three patterns are
consistent across assets: high volatility persistence in both sessions,
asymmetric leverage effects that vary across sessions and sectors, and
cross-session spillovers with a visible sectoral pattern.

\begin{table}[!t]
\centering
\caption{Mean Equation Estimates}
\label{tab:Mean-Equation}
\vspace{0.2em}
\begin{threeparttable}
	\footnotesize
	\setlength{\tabcolsep}{2pt}
	\begin{tabularx}{\textwidth}{l YYYYYYY}
		\toprule
        \midrule
		& \multicolumn{7}{c}{Parameter Estimates} \\
		\cmidrule(lr){2-8}
		Asset & $\mu_N$ & $\mu_D$ & $\delta$ & $\phi_{11}$ & $\phi_{12}$ & $\phi_{21}$ & $\phi_{22}$ \\
		\midrule
		\textbf{Energy Sector} \\
		\midrule
		CVX & 0.0227 & -0.0058 & 0.0636 & -0.0223 & 0.0083 & 0.0304 & -0.0354 \\
		& (0.0096) & (0.0154) & (0.0227) & (0.0152) & (0.0099) & (0.0223) & (0.0152) \\
		\addlinespace[0.5ex]
		APA & 0.0357 & -0.0755 & 0.0162 & -0.0104 & 0.0240 & 0.0265 & -0.0029 \\
		& (0.0151) & (0.0260) & (0.0309) & (0.0142) & (0.0092) & (0.0251) & (0.0155) \\
		\addlinespace[0.5ex]
		DVN & 0.0572 & -0.0641 & 0.0134 & -0.0098 & 0.0231 & 0.0165 & -0.0259 \\
		& (0.0140) & (0.0255) & (0.0266) & (0.0148) & (0.0090) & (0.0251) & (0.0147) \\
		\addlinespace[1.5ex]
		\midrule
		\textbf{Technology Sector} \\
		\midrule
		MSFT & 0.0228 & 0.0413 & -0.0136 & 0.0048 & -0.0461 & -0.0028 & -0.0309 \\
		& (0.0127) & (0.0150) & (0.0240) & (0.0150) & (0.0132) & (0.0226) & (0.0145) \\
		\addlinespace[0.5ex]
		INTC & 0.0017 & 0.0093 & -0.0029 & 0.0017 & -0.0311 & 0.0631 & -0.0081 \\
		& (0.0150) & (0.0182) & (0.0232) & (0.0156) & (0.0148) & (0.0189) & (0.0155) \\
		\addlinespace[0.5ex]
		CSCO & -0.0065 & 0.0396 & -0.0363 & -0.0046 & -0.0284 & 0.0721 & 0.0040 \\
		& (0.0142) & (0.0151) & (0.0221) & (0.0125) & (0.0143) & (0.0174) & (0.0165) \\
		\midrule
		\bottomrule
	\end{tabularx}
	\begin{tablenotes}[flushleft]
		\footnotesize
		\item \textit{Note}: This table reports the parameter estimates for the
conditional mean equation. Parzen-kernel HAC standard errors, computed using the
estimated conditional variances, are reported in parentheses.
	\end{tablenotes}
\end{threeparttable}
\vspace{-0.5em}
\end{table}

Table \ref{tab:Mean-Equation} reports the mean equation estimates. The
parameters $\mu_N$ and $\mu_D$ are intercepts in the overnight and intraday
mean equations, respectively. Energy stocks tend to exhibit positive overnight
intercepts and negative intraday intercepts, whereas technology stocks show
heterogeneous overnight intercepts and uniformly positive intraday intercepts.
This suggests that return dynamics differ systematically across sectors,
particularly in how returns are distributed across trading and non-trading
hours.

The point estimates of the contemporaneous transmission parameter $\delta_i$
suggest a sectoral contrast. As implied by Equation~\ref{eq:EQ1}, the intraday
return $R_{i,t}^{D}$ depends on the term $\delta_i R_{i,t}^{N}$. Accordingly,
positive estimates of $\delta_i$ indicate overnight--intraday continuation,
whereas negative estimates indicate reversal. The estimates are positive for
the energy stocks and negative for the technology stocks, although the
statistical strength varies across assets. This suggests that overnight
information in energy stocks tends to be reinforced during the trading day,
while technology stocks exhibit partial correction of overnight price movements.
Nevertheless, the magnitude of $\delta_i$ remains modest across both sectors.

The autoregressive coefficients in $\Phi_i$ are small across all assets. While
significance varies, the economically small magnitudes suggest that lagged
return dynamics play a limited role, whereas the contemporaneous transmission
parameter $\delta_i$ captures the more visible within-day relation between
overnight and intraday returns.

Table \ref{tab:Volatility-Dynamics} presents the volatility equation estimates,
divided into Panel A (Overnight Volatility) and Panel B (Intraday Volatility).
Volatility persistence is high in both sessions. The autoregressive coefficients
$\beta^{N}$ and $\beta^{D}$ are close to unity, indicating strong volatility
clustering in both trading periods.

Leverage effects differ across sessions. Intraday volatility displays a
standard leverage effect, with negative shocks generating larger increases in
volatility. Overnight volatility shows more heterogeneity. Energy stocks retain
the conventional leverage pattern, while some technology stocks show weaker or
even reversed asymmetry during non-trading hours.

Cross-session spillovers are economically meaningful for many assets,
especially through the magnitude spillover terms. The magnitude spillover
parameters $\delta_{2}^{N}$ and $\delta_{2}^{D}$ are uniformly positive,
indicating that large shocks in one session lead to increased volatility in the
next. The signed spillover parameters $\delta_{1}^{N}$ and $\delta_{1}^{D}$
display more heterogeneous patterns. For energy stocks, $\delta_{1}^{N}$ is
consistently negative across all three firms, suggesting that positive intraday
shocks are associated with reduced overnight volatility. For technology stocks,
$\delta_{1}^{N}$ is mixed in sign, offering little directional information. The
pattern for $\delta_{1}^{D}$ is similarly heterogeneous across sectors, with
most estimates negative but varying in magnitude and significance.

\begin{table}[!t]
\centering
\caption{Volatility Equation Estimates}
\label{tab:Volatility-Dynamics}
\vspace{0.2em}
\begin{threeparttable}
	\footnotesize
	\setlength{\tabcolsep}{2pt}
	\renewcommand{\arraystretch}{0.78}
	
	\begin{tabularx}{\textwidth}{l YYYYYY}
		\toprule
		\midrule
		& \multicolumn{3}{c}{\textbf{Energy Sector}} 
		& \multicolumn{3}{c}{\textbf{Technology Sector}} \\
		\cmidrule(lr){2-4} \cmidrule(lr){5-7}
		& CVX & APA & DVN & MSFT & INTC & CSCO \\
		\midrule

		\multicolumn{7}{l}{\textbf{Panel A: Overnight Volatility}} \\
		\midrule
		$\omega^{N}$ & -0.1890 & -0.1640 & -0.1803 & -0.0943 & -0.1528 & -0.1774 \\
		& (0.0330) & (0.0463) & (0.0314) & (0.0491) & (0.0803) & (0.0579) \\
		$\beta^{N}$ & 0.9851 & 0.9902 & 0.9864 & 0.9769 & 0.9768 & 0.9426 \\
		& (0.0099) & (0.0038) & (0.0039) & (0.0092) & (0.0193) & (0.0225) \\
		$\tau_{1}^{N}$ & -0.0601 & -0.0664 & -0.0562 & -0.0473 & 0.0577 & 0.0056 \\
		& (0.0145) & (0.0212) & (0.0152) & (0.0190) & (0.0297) & (0.0573) \\
		$\tau_{2}^{N}$ & 0.1316 & 0.1126 & 0.1071 & 0.0675 & 0.0675 & 0.0263 \\
		& (0.0272) & (0.0403) & (0.0256) & (0.0228) & (0.0313) & (0.0333) \\
		$\delta_{1}^{N}$ & -0.0523 & -0.0357 & -0.0392 & 0.0101 & -0.0324 & -0.0389 \\
		& (0.0211) & (0.0189) & (0.0122) & (0.0141) & (0.0308) & (0.0392) \\
		$\delta_{2}^{N}$ & 0.1080 & 0.1162 & 0.1375 & 0.0653 & 0.1544 & 0.2207 \\
		& (0.0206) & (0.0319) & (0.0273) & (0.0550) & (0.0915) & (0.0775) \\

		\addlinespace[0.3ex]
		\midrule
		\multicolumn{7}{l}{\textbf{Panel B: Intraday Volatility}} \\
		\midrule
		$\omega^{D}$ & -0.1800 & -0.1213 & -0.1263 & -0.1555 & -0.1701 & -0.1619 \\
		& (0.0303) & (0.0303) & (0.0189) & (0.0249) & (0.0380) & (0.0286) \\
		$\beta^{D}$ & 0.9799 & 0.9894 & 0.9858 & 0.9692 & 0.9736 & 0.9662 \\
		& (0.0119) & (0.0036) & (0.0038) & (0.0107) & (0.0109) & (0.0094) \\
		$\tau_{1}^{D}$ & -0.0503 & -0.0365 & -0.0304 & -0.0215 & -0.0223 & -0.0005 \\
		& (0.0289) & (0.0127) & (0.0077) & (0.0202) & (0.0162) & (0.0158) \\
		$\tau_{2}^{D}$ & 0.1506 & 0.1109 & 0.1198 & 0.1308 & 0.1767 & 0.1856 \\
		& (0.0169) & (0.0289) & (0.0201) & (0.0273) & (0.0432) & (0.0264) \\
		$\delta_{1}^{D}$ & -0.0555 & -0.0425 & -0.0335 & -0.0232 & 0.0027 & -0.0476 \\
		& (0.0175) & (0.0124) & (0.0086) & (0.0123) & (0.0197) & (0.0263) \\
		$\delta_{2}^{D}$ & 0.0929 & 0.0723 & 0.0715 & 0.1068 & 0.0847 & 0.0519 \\
		& (0.0268) & (0.0215) & (0.0151) & (0.0210) & (0.0262) & (0.0260) \\
		\midrule
		\bottomrule
	\end{tabularx}

	\begin{tablenotes}[flushleft]
		\scriptsize
		\setlength{\itemsep}{0pt}
		\item \textit{Note}: This table reports the volatility equation estimates. Panel A presents overnight volatility parameters; Panel B presents intraday volatility parameters. Robust standard errors are in parentheses.
	\end{tablenotes}
\end{threeparttable}
\vspace{-0.5em}
\end{table}

\subsection{Multivariate Model Estimates\label{subsec:Six-Asset-Multivariate}}

The multivariate model results are organized around two dimensions, the
structure of which is illustrated in Figure \ref{fig:Return-Decomposition-and-1}.
The first is the tail partition: a session-level specification where overnight
and intraday innovations each share a single degrees-of-freedom parameter
across all assets; a session-by-sector specification where Energy and
Technology assets carry distinct tail parameters within each session; and an
asset-level specification where each asset in each session has its own
degrees-of-freedom parameter. The second dimension is whether a sector-based
block structure is imposed on the correlation matrix. As an additional
benchmark, we also consider a clustering based purely on sector classification,
which does not distinguish between overnight and intraday sessions. The results,
reported in Appendix~\ref{sec:Sector-Based-Clustering}, show that this
specification performs worse than the session-based partitions, highlighting the
importance of session-level heterogeneity in shaping tail behavior.

Tables \ref{tab:DCC_GAS} and \ref{tab:Estimations-of-Score-driven} present
results without the block structure, covering the full range of tail partitions
within the score-driven framework alongside the DCC benchmark. Table
\ref{tab:Estimations-of-Block-1} then imposes the block structure across all
four tail specifications.

\subsubsection{Baseline comparison}

Table \ref{tab:DCC_GAS} compares the DCC model introduced in Section \ref{sec:Benchmark-DCC} with the
score-driven framework under a common set of distributional assumptions. Both
models are estimated with an unrestricted correlation matrix. For the
heavy-tailed specifications, we consider both a common multivariate-$t$
distribution and a session-level Cluster-$t$ distribution, where tail thickness
is allowed to differ between overnight and intraday innovations. We also report
the Gaussian case as a limiting benchmark. 

\begin{table}[!t]
\centering
\caption{Score-Driven and DCC Model Comparison}
\label{tab:DCC_GAS}
\vspace{0.2em}
\begin{threeparttable}
	\footnotesize
	\setlength{\tabcolsep}{2.5pt}
	\renewcommand{\arraystretch}{0.84}
	
	\begin{tabularx}{\textwidth}{l c YYYYYY}
		\toprule
		\midrule
		& & \multicolumn{2}{c}{Cluster-$t$ (Session, $G=2$)} 
		  & \multicolumn{2}{c}{Multivariate-$t$ ($G=1$)} 
		  & \multicolumn{2}{c}{Gaussian} \\
		\cmidrule(lr){3-4} \cmidrule(lr){5-6} \cmidrule(lr){7-8}
		& & Night & Day & Night & Day & Night & Day \\
		\midrule
		\multicolumn{8}{l}{\textbf{Panel A: DCC Model}} \\
		\midrule
		\multirow{3}{*}{$\mu$}
		& $Q_{25}$ & 0.1612 & 0.1275 & 0.2252 & 0.1054 & 0.0842 & 0.1202 \\
		& $Q_{50}$ & 0.2022 & 0.1719 & 0.2824 & 0.1352 & 0.1320 & 0.1682 \\
		& Max      & 0.7970 & 0.7945 & 0.9890 & 0.6846 & 0.5582 & 0.7562 \\
		\addlinespace[0.2ex]
		\multirow{3}{*}{$\beta$}
		& $Q_{25}$ & 0.9861 & 0.9643 & 0.9690 & 0.9628 & 0.8595 & 0.9611 \\
		& $Q_{50}$ & 0.9875 & 0.9723 & 0.9792 & 0.9688 & 0.8779 & 0.9684 \\
		& Max      & 0.9928 & 0.9803 & 0.9942 & 0.9809 & 0.9284 & 0.9755 \\
		\addlinespace[0.2ex]
		\multirow{3}{*}{$\alpha$}
		& $Q_{25}$ & 0.0032 & 0.0113 & 0.0019 & 0.0122 & 0.0603 & 0.0138 \\
		& $Q_{50}$ & 0.0061 & 0.0141 & 0.0053 & 0.0137 & 0.0687 & 0.0171 \\
		& Max      & 0.0150 & 0.0258 & 0.0175 & 0.0226 & 0.1314 & 0.0247 \\
		\addlinespace[0.2ex]
	    $\nu$  && 3.2336 & 10.791 & \multicolumn{2}{c}{5.6031} & \multicolumn{2}{c}{} \\
		\addlinespace[0.2ex]
	    $p$      & & \multicolumn{2}{c}{116} & \multicolumn{2}{c}{115} & \multicolumn{2}{c}{114} \\
        $\ell$   & & \multicolumn{2}{c}{\textbf{-65472}} & \multicolumn{2}{c}{-67281} & \multicolumn{2}{c}{-73937} \\
        $\ell_N$ & & \multicolumn{2}{c}{\textbf{-28955}} & \multicolumn{2}{c}{-29512} & \multicolumn{2}{c}{-36935} \\
        $\ell_D$ & & \multicolumn{2}{c}{\textbf{-36516}} & \multicolumn{2}{c}{-37769} & \multicolumn{2}{c}{-37002} \\
        BIC      & & \multicolumn{2}{c}{\textbf{131932}} & \multicolumn{2}{c}{135542} & \multicolumn{2}{c}{148846} \\
\addlinespace[0.3ex]
\midrule
\multicolumn{8}{l}{\textbf{Panel B: Score-driven Model}} \\
	\midrule
		\multirow{3}{*}{$\mu$}
		& $Q_{25}$ & 0.1660 & 0.1284 & 0.2001 & 0.1092 & 0.0575 & 0.1265 \\
		& $Q_{50}$ & 0.1970 & 0.1736 & 0.2591 & 0.1474 & 0.1445 & 0.1681 \\
		& Max      & 0.8313 & 0.7923 & 0.8951 & 0.7462 & 0.6486 & 0.7817 \\
		\addlinespace[0.5ex]
		\multirow{3}{*}{$\beta$}
		& $Q_{25}$ & 0.9910 & 0.9764 & 0.9403 & 0.9586 & 0.9002 & 0.9763 \\
		& $Q_{50}$ & 0.9950 & 0.9860 & 0.9946 & 0.9864 & 0.9892 & 0.9843 \\
		& Max      & 0.9995 & 0.9975 & 0.9995 & 0.9976 & 0.9986 & 0.9983 \\
		\addlinespace[0.5ex]
		\multirow{3}{*}{$\alpha$}
		& $Q_{25}$ & 0.0065 & 0.0073 & 0.0054 & 0.0065 & 0.0050 & 0.0058 \\
		& $Q_{50}$ & 0.0111 & 0.0107 & 0.0119 & 0.0114 & 0.0074 & 0.0110 \\
		& Max      & 0.0608 & 0.0330 & 0.0747 & 0.0218 & 0.0932 & 0.0376 \\
		\addlinespace[0.5ex]
		$\nu$ && 3.3305 & 10.958 & \multicolumn{2}{c}{5.6488} & \multicolumn{2}{c}{} \\
		\addlinespace[0.5ex]
		$p$      & & \multicolumn{2}{c}{92} & \multicolumn{2}{c}{91} & \multicolumn{2}{c}{90} \\
        $\ell$   & & \multicolumn{2}{c}{\textbf{-65242}} & \multicolumn{2}{c}{-67051} & \multicolumn{2}{c}{-74102} \\
        $\ell_N$ & & \multicolumn{2}{c}{\textbf{-28739}} & \multicolumn{2}{c}{-29284} & \multicolumn{2}{c}{-37110} \\
        $\ell_D$ & & \multicolumn{2}{c}{\textbf{-36503}} & \multicolumn{2}{c}{-37767} & \multicolumn{2}{c}{-36992} \\
        BIC      & & \multicolumn{2}{c}{\textbf{131268}} & \multicolumn{2}{c}{134878} & \multicolumn{2}{c}{148971} \\
		\midrule
		\bottomrule
	\end{tabularx}
		\begin{tablenotes}[flushleft]
		\scriptsize
		\setlength{\itemsep}{0pt}
		\item \textit{Note}: This table compares the DCC and score-driven models under the Cluster-$t$,
multivariate-$t$, and Gaussian distributions, estimated without block structure.
For DCC models, $\mu$ is the unconditional correlation parameter; for
score-driven models, $\mu$ is the mean parameter in the transformed correlation
space. The parameters $\beta$ and $\alpha$ denote persistence and score
sensitivity, respectively. $Q_{25}$ and $Q_{50}$ denote the 25th and 50th
percentiles. $\ell_N$ and $\ell_D$ are the decomposed log-likelihoods for the
overnight and intraday sessions. Gaussian specifications do not have a
degrees-of-freedom parameter $\nu$.  The parameter count $p$ includes only second-stage dependence and tail
parameters used in the likelihood comparison and BIC calculation; first-stage
Coupled EGARCH parameters are excluded. The larger $p$ for the DCC
specifications reflects the additional unconditional correlation parameters in
the DCC recursion. Bold values indicate, within each panel (DCC or score-driven), the specification with the highest log-likelihood.
	\end{tablenotes}
\end{threeparttable}
\vspace{-0.6em}
\end{table}

The score-driven framework consistently outperforms the DCC model under
heavy-tailed specifications, while using fewer parameters. Under the
Cluster-$t$, the total log-likelihood increases by 230 points, with 216 points
coming from the overnight session and only 13 points from the intraday session.
A similar pattern holds under the multivariate-$t$ case. This asymmetry is
consistent with the heavier tails of overnight returns, where extreme
observations are more frequent and their treatment matters more for correlation
updating.

Under the Gaussian specification, however, the ordering reverses for the
overnight session: the score-driven model exhibits a deterioration of about 175
log-likelihood points, while the intraday component remains broadly comparable.
When $W_{t}=1$, the Gaussian score is unbounded, so extreme overnight shocks
enter the correlation recursion without any down-weighting. This reversal under
the Gaussian specification suggests that the score-driven advantage depends on
an appropriate heavy-tailed specification, and further motivates the use of
heavy-tailed distributions for overnight returns.

Comparing across distributions, moving from the Gaussian benchmark to the
multivariate-$t$ specification leads to large log-likelihood gains exceeding
6,000 points in both the DCC and score-driven models. These gains are entirely
driven by the overnight component: the overnight log-likelihood improves by more
than 7,000 points, whereas the intraday component deteriorates under the common
multivariate-$t$ specification. This pattern reflects a limitation of imposing a
single degrees-of-freedom parameter across sessions. Because overnight returns
are substantially more heavy-tailed, the common tail parameter is largely driven
by the overnight distribution. The resulting tail thickness is too heavy for
intraday returns, leading to a poorer fit for that component. Thus, the common
multivariate-$t$ restriction forces a compromise that cannot fully capture the
distinct tail behavior of the two sessions.

Allowing for tail heterogeneity across sessions directly addresses this
limitation. Relative to the multivariate-$t$ benchmark, the Cluster-$t$
specification delivers sizable gains for both sessions. In the score-driven
model, the overnight log-likelihood improves by about 545 points, while the
intraday component improves by about 1,264 points. A similar pattern is observed
under DCC. The larger improvement for the intraday component reflects the
correction of the excessively heavy tail imposed by the common multivariate-$t$
specification. The estimated degrees of freedom reveal pronounced heterogeneity,
with much heavier tails for overnight returns ($\nu_{N}=3.3305$) than intraday
returns ($\nu_{D}=10.9582$), indicating that a single tail parameter is
insufficient.

Finally, these gains are achieved without sacrificing parsimony. The
score-driven Cluster-$t$ model has fewer parameters than the DCC Cluster-$t$
model and improves the BIC by about 664 points. Relative to the score-driven
multivariate-$t$ specification, it improves the BIC by about 3,610 points,
confirming that the additional session-specific tail parameter is strongly
supported by the information criterion.

\subsubsection{Tail heterogeneity by session, sector, and asset}

Table \ref{tab:Estimations-of-Score-driven} presents three tail partition
specifications within the score-driven framework. Score-Cluster-$t$ (Session,
$G=2$) assigns a single degrees-of-freedom parameter to each session.
Score-Cluster-$t$ (Session $\times$ Sector, $G=4$) further partitions each
session into Energy and Technology clusters. Score-Hetero-$t$ (Session
$\times$ Asset, $G=12$) allows for asset-level tail parameters. The correlation
matrix remains unrestricted across all three cases.

\begin{table}[!t]
\centering
\caption{Score-Driven Models with Alternative Tail Partitions 
\label{tab:Estimations-of-Score-driven}}
\vspace{0.2em}
\begin{threeparttable}
	\footnotesize
	\setlength{\tabcolsep}{2.5pt}
	\renewcommand{\arraystretch}{0.86}
	
	\begin{tabularx}{\textwidth}{l c YYYYYY}
		\toprule
		\midrule
        & & \multicolumn{2}{c}{Score-Cluster-$t$} 
          & \multicolumn{2}{c}{Score-Cluster-$t$} 
          & \multicolumn{2}{c}{Score-Hetero-$t$} \\
        & & \multicolumn{2}{c}{(Session, $G=2$)} 
          & \multicolumn{2}{c}{(Session $\times$ Sector, $G=4$)} 
          & \multicolumn{2}{c}{(Session $\times$ Asset, $G=12$)} \\
        \cmidrule(lr){3-4} \cmidrule(lr){5-6} \cmidrule(lr){7-8}
        & & Night & Day & Night & Day & Night & Day \\
		\midrule
		
		\multicolumn{8}{l}{\textbf{Panel A: Parameter Estimates}} \\
			\midrule
		\multirow{3}{*}{$\mu$}
		& $Q_{25}$ & 0.1660 & 0.1284 & 0.1548 & 0.1270 & 0.1630 & 0.1302 \\
        & $Q_{50}$ & 0.1970 & 0.1736 & 0.1937 & 0.1728 & 0.2123 & 0.1780 \\
		& Max      & 0.8313 & 0.7923 & 0.9342 & 0.7967 & 0.8844 & 0.7897 \\
		
		\addlinespace[0.2ex]
		\multirow{3}{*}{$\beta$}
		& $Q_{25}$ & 0.9910 & 0.9764 & 0.9754 & 0.9783 & 0.9849 & 0.9809 \\
        & $Q_{50}$ & 0.9950 & 0.9860 & 0.9909 & 0.9862 & 0.9921 & 0.9842 \\
		& Max      & 0.9995 & 0.9975 & 0.9997 & 0.9976 & 0.9993 & 0.9976 \\
		
		\addlinespace[0.2ex]
		\multirow{3}{*}{$\alpha$}
		& $Q_{25}$ & 0.0065 & 0.0073 & 0.0097 & 0.0077 & 0.0092 & 0.0074 \\
        & $Q_{50}$ & 0.0111 & 0.0107 & 0.0148 & 0.0114 & 0.0133 & 0.0104 \\
		& Max      & 0.0608 & 0.0330 & 0.0416 & 0.0325 & 0.1114 & 0.0320 \\
		
		\addlinespace[0.3ex]
        \midrule
        \multicolumn{8}{l}{\textbf{Panel B: Degrees of Freedom}} \\
        	\midrule
        $\nu_0$ && 3.3305 & 10.958 & & & & \\
        $\nu_1$ &&        &         & \multirow{3}{*}{3.7859} & \multirow{3}{*}{8.8750} & 3.8579 & 8.3447 \\
        $\nu_2$ &&        &         &                         &                         & 3.1788 & 6.5741 \\
        $\nu_3$ &&        &         &                         &                         & 3.4367 & 8.2856 \\
        $\nu_4$ &&        &         & \multirow{3}{*}{2.6478} & \multirow{3}{*}{8.7139} & 2.6441 & 7.4232 \\
        $\nu_5$ &&        &         &                         &                         & 2.5225 & 7.1219 \\
        $\nu_6$ &&        &         &                         &                         & 2.3954 & 7.4850 \\
        
        \addlinespace[0.3ex]
		\midrule
		\multicolumn{8}{l}{\textbf{Panel C: Model Fit Diagnostics}} \\
			\midrule
		$p$       && \multicolumn{2}{c}{92}      & \multicolumn{2}{c}{94}             & \multicolumn{2}{c}{102} \\
		$\ell$   && \multicolumn{2}{c}{-65242}  & \multicolumn{2}{c}{\textbf{-64071}} & \multicolumn{2}{c}{-64350} \\
		$\ell_N$ && \multicolumn{2}{c}{-28739}  & \multicolumn{2}{c}{\textbf{-27657}} & \multicolumn{2}{c}{-27934} \\
		$\ell_D$ && \multicolumn{2}{c}{-36503}  & \multicolumn{2}{c}{\textbf{-36414}} & \multicolumn{2}{c}{-36416} \\
		BIC      && \multicolumn{2}{c}{131268}  & \multicolumn{2}{c}{\textbf{128943}} & \multicolumn{2}{c}{129569} \\
		\midrule
		\bottomrule
	\end{tabularx}
	
	\begin{tablenotes}[flushleft]
		\scriptsize
		\setlength{\itemsep}{0pt}
		\item \textit{Note}: This table reports score-driven model estimates under three tail specifications without block structure. Score-Cluster-$t$ (Session, $G=2$) assigns one degrees-of-freedom parameter to each session; Score-Cluster-$t$ (Session $\times$ Sector, $G=4$) partitions each session into Energy and Technology clusters; and Score-Hetero-$t$ (Session $\times$ Asset, $G=12$) assigns each asset within each session its own degrees-of-freedom parameter. $Q_{25}$ and $Q_{50}$ denote the 25th and 50th percentiles. Under the session-level Cluster-$t$, $\nu_0$ denotes the session-specific degrees of freedom. Under the session-by-sector Cluster-$t$, $\nu_1$--$\nu_3$ share the Energy sector parameter and $\nu_4$--$\nu_6$ share the Technology sector parameter. Under Hetero-$t$ ($G=12$), in each session column, $\nu_1, \ldots, \nu_6$ correspond to CVX, APA, DVN, MSFT, INTC, and CSCO. $\ell_N$ and $\ell_D$ are decomposed log-likelihoods. The parameter count $p$ includes only second-stage dependence and tail
parameters used in the likelihood comparison and BIC calculation. Bold values indicate the superior specification.
	\end{tablenotes}
\end{threeparttable}
\vspace{-0.6em}
\end{table}

Panel A reports dynamic parameters across specifications. The
persistence $\beta$ is uniformly high under all specifications, and the score
sensitivity $\alpha$ shows similar stability. Panel B highlights a clear and
persistent difference between overnight and intraday returns. Under the
session-level specification, the overnight degrees of freedom are around 3.3,
compared to roughly 11 for intraday returns. This gap remains when the
partition is refined. At the sector level, overnight tails range from about 2.6
to 3.8, while intraday values cluster tightly around 9. The same separation
appears in the Hetero-$t$ specification: overnight degrees of freedom lie
between 2.4 and 3.9, whereas intraday values fall between 6.6 and 8.3, with
little overlap between the two. Regardless of how the partition is defined,
overnight returns consistently exhibit much heavier tails. There is also some
heterogeneity within the overnight session. Technology stocks display heavier
tails than Energy stocks, with degrees of freedom around 2.6 versus 3.8 under
the sector-level specification. This difference is consistent with the greater
exposure of technology firms to after-hours information releases. During the
trading day, however, the two sectors look very similar, with degrees of freedom
close to 9 in both cases.

In terms of model fit, the session-by-sector Cluster-$t$ provides the best
performance and lowest BIC. It improves the log-likelihood by 1,171 points relative to the
session-level specification and by 279 points relative to the Hetero-$t$. The improvement over the session-level
specification is mainly driven by the overnight component: moving from the
session-level to the session-by-sector specification increases the overnight
log-likelihood by about 1,082 points, while the intraday contribution is only 89
points. The weaker performance of the Hetero-$t$ suggests that, in this
six-asset application, the additional marginal tail flexibility does not
compensate for the loss of a shared sector-level tail component. The
session-by-sector Cluster-$t$ provides a better balance between tail
flexibility and dependence modeling.

\subsubsection{Block correlation structure and dynamic correlations}

Table \ref{tab:Estimations-of-Block-1} imposes a sector-based block structure
across all four tail specifications for comparison with the unrestricted
results in Tables \ref{tab:DCC_GAS} and \ref{tab:Estimations-of-Score-driven}.

\begin{table}[!t]
\centering
\caption{Block Correlation Structure Estimates 
\label{tab:Estimations-of-Block-1}}
\vspace{0.2em}
\begin{threeparttable}
\footnotesize
\setlength{\tabcolsep}{2.2pt}
\renewcommand{\arraystretch}{0.82}
	
	\begin{tabularx}{\textwidth}{l c YYYYYYYY}
		\toprule
		\midrule
		& & \multicolumn{2}{c}{Score-Block-Multivariate-$t$} 
		& \multicolumn{2}{c}{Score-Block-Cluster-$t$} 
		& \multicolumn{2}{c}{Score-Block-Cluster-$t$} 
		& \multicolumn{2}{c}{Score-Block-Hetero-$t$} \\
		
		& & \multicolumn{2}{c}{}
		& \multicolumn{2}{c}{Session}
		& \multicolumn{2}{c}{Session $\times$ Sector}
		& \multicolumn{2}{c}{Session $\times$ Asset} \\
		& & \multicolumn{2}{c}{($G=1$)}
		& \multicolumn{2}{c}{($G=2$)}
		& \multicolumn{2}{c}{($G=4$)}
		& \multicolumn{2}{c}{($G=12$)} \\

		\cmidrule(lr){3-4} \cmidrule(lr){5-6} \cmidrule(lr){7-8} \cmidrule(lr){9-10}
		& & Night & Day & Night & Day & Night & Day & Night & Day \\
		\midrule
		\multicolumn{10}{l}{\textbf{Panel A: Implied Correlations and Parameter Estimates}} \\
		\midrule
		\multirow{3}{*}{$\rho$}
		& $\rho_{11}$ & 0.7805 & 0.6244 & 0.7341 & 0.6667 & 0.7827 & 0.6697 & 0.7650 & 0.6670 \\
		& $\rho_{12}$ & 0.4539 & 0.2266 & 0.3821 & 0.2724 & 0.3543 & 0.2710 & 0.3475 & 0.2730 \\
		& $\rho_{22}$ & 0.6957 & 0.4464 & 0.6447 & 0.4948 & 0.5895 & 0.4971 & 0.5102 & 0.4957 \\
		\addlinespace[0.2ex]
		\multirow{3}{*}{$\mu$}
		& $\mu_{11}$ & 0.7580 & 0.5724 & 0.6967 & 0.6209 & 0.7979 & 0.6258 & 0.7648 & 0.6218 \\
		& $\mu_{12}$ & 0.2018 & 0.1097 & 0.1721 & 0.1294 & 0.1595 & 0.1285 & 0.1590 & 0.1285 \\
		& $\mu_{22}$ & 0.6151 & 0.3822 & 0.5662 & 0.4210 & 0.5108 & 0.4253 & 0.4204 & 0.4238 \\
		\addlinespace[0.2ex]
		\multirow{3}{*}{$\beta$}
		& $\beta_{11}$ & 0.9929 & 0.9893 & 0.9919 & 0.9842 & 0.9876 & 0.9816 & 0.9864 & 0.9829 \\
		& $\beta_{12}$ & 0.9735 & 0.9864 & 0.9814 & 0.9830 & 0.9789 & 0.9820 & 0.9811 & 0.9829 \\
		& $\beta_{22}$ & 0.9138 & 0.9782 & 0.9349 & 0.9827 & 0.9603 & 0.9800 & 0.9332 & 0.9809 \\
		\addlinespace[0.2ex]
		\multirow{3}{*}{$\alpha$}
		& $\alpha_{11}$ & 0.0293 & 0.0255 & 0.0338 & 0.0318 & 0.0452 & 0.0340 & 0.0454 & 0.0336 \\
		& $\alpha_{12}$ & 0.0891 & 0.0236 & 0.0624 & 0.0388 & 0.0792 & 0.0407 & 0.0685 & 0.0394 \\
		& $\alpha_{22}$ & 0.0531 & 0.0322 & 0.0549 & 0.0348 & 0.0454 & 0.0393 & 0.0787 & 0.0387 \\
		\addlinespace[0.3ex]
        \midrule
         \multicolumn{10}{l}{\textbf{Panel B: Degrees of Freedom}} \\
        	\midrule
		$\nu_0$ 
		&& \multicolumn{2}{c}{5.6532} & 3.3450 & 10.951 & & & & \\
		$\nu_1$ 
		&& \multicolumn{2}{c}{} & & & \multirow{3}{*}{3.8109} & \multirow{3}{*}{8.7091} & 3.9607 & 9.0622 \\
		$\nu_2$ 
		&& \multicolumn{2}{c}{} & & & & & 3.1909 & 6.3936 \\
		$\nu_3$
		&& \multicolumn{2}{c}{} & & & & & 3.3801 & 7.9721 \\
		$\nu_4$ 
		&& \multicolumn{2}{c}{} & & & \multirow{3}{*}{2.6515} & \multirow{3}{*}{8.8394} & 2.6281 & 7.7632 \\
		$\nu_5$
		&& \multicolumn{2}{c}{} & & & & & 2.5188 & 7.2765 \\
		$\nu_6$
		&& \multicolumn{2}{c}{} & & & & & 2.4063 & 7.2252 \\
		
		\addlinespace[0.3ex]
        \midrule
        \multicolumn{10}{l}{\textbf{Panel C: Model Fit Diagnostics}} \\
        	\midrule
		$p$
		&& \multicolumn{2}{c}{19}
		& \multicolumn{2}{c}{20}
		& \multicolumn{2}{c}{\textbf{22}}
		& \multicolumn{2}{c}{30} \\
		$\ell$
		&& \multicolumn{2}{c}{-67128}
		& \multicolumn{2}{c}{-65337}
		& \multicolumn{2}{c}{\textbf{-64139}}
		& \multicolumn{2}{c}{-64438} \\
		$\ell_N$
		&& \multicolumn{2}{c}{-29251}
		& \multicolumn{2}{c}{-28727}
		& \multicolumn{2}{c}{\textbf{-27627}}
		& \multicolumn{2}{c}{-27909} \\
		$\ell_D$
		&& \multicolumn{2}{c}{-37877}
		& \multicolumn{2}{c}{-36610}
		& \multicolumn{2}{c}{\textbf{-36512}}
		& \multicolumn{2}{c}{-36529} \\
		BIC
		&& \multicolumn{2}{c}{134417}
		& \multicolumn{2}{c}{130845}
		& \multicolumn{2}{c}{\textbf{128465}}
		& \multicolumn{2}{c}{129132} \\
		\midrule
		\bottomrule
	\end{tabularx}
	\begin{tablenotes}[flushleft]
	\scriptsize
	\setlength{\itemsep}{0pt}
	\item \textit{Note}: This table reports score-driven model estimates under four specifications with a sector-based block structure imposed on the correlation matrix. $\rho_{11}$, $\rho_{22}$, and $\rho_{12}$ denote implied unconditional correlations for intra-Energy, intra-Technology, and cross-sector pairs, respectively. $\mu$, $\beta$, and $\alpha$ denote the corresponding condensed parameters, persistence parameters, and score sensitivity parameters. $\nu_0$ denotes the global or session-specific degrees of freedom for models without cross-sectional tail clusters ($G=1,2$). Under the sector-based Cluster-$t$ model ($G=4$), $\nu_1$--$\nu_3$ share the Energy-sector parameter and $\nu_4$--$\nu_6$ share the Technology-sector parameter. Under Hetero-$t$ ($G=12$), in each session column, $\nu_1, \ldots, \nu_6$ correspond to CVX, APA, DVN, MSFT, INTC, and CSCO. $\ell_N$ and $\ell_D$ are decomposed log-likelihoods.
    The parameter count $p$ includes only second-stage dependence and tail
parameters used in the likelihood comparison and BIC calculation. Bold values indicate the superior specification.
\end{tablenotes}
\end{threeparttable}
\vspace{-0.6em}
\end{table}

The results are closely aligned with those obtained under the unrestricted
specifications. The multivariate-$t$ provides a relatively poor fit, with a
log-likelihood of $-67,128$ and an implied degrees-of-freedom of about 5.7.
Allowing for session-specific tails increases the log-likelihood by about 1,791
points with only one additional tail parameter, and improves the BIC by about
3,572 points.

The ranking of tail partitions is unchanged after imposing the block structure.
The session-by-sector Cluster-$t$ again provides the best fit, improving the
log-likelihood by 1,198 points relative to the session-level specification, with
the gain concentrated in the overnight component. The fully flexible Hetero-$t$
does not improve fit: its log-likelihood is 299 points lower than that of the
session-by-sector Cluster-$t$, despite using more parameters.

Across all specifications, the block structure sharply reduces the number of
parameters. Although the raw log-likelihood declines slightly, the loss is small
relative to the large reduction in dimensionality, leading to improved BIC
values. For the session-level Cluster-$t$, parameters fall from 92 to 20, and
the BIC improves by 423 points. The session-by-sector specification yields an
even larger BIC gain of 478, while the Hetero-$t$ also benefits from a
reduction in complexity, with a BIC improvement of 437 points.

\begin{figure}
\centering
\includegraphics[width=0.9\textwidth]{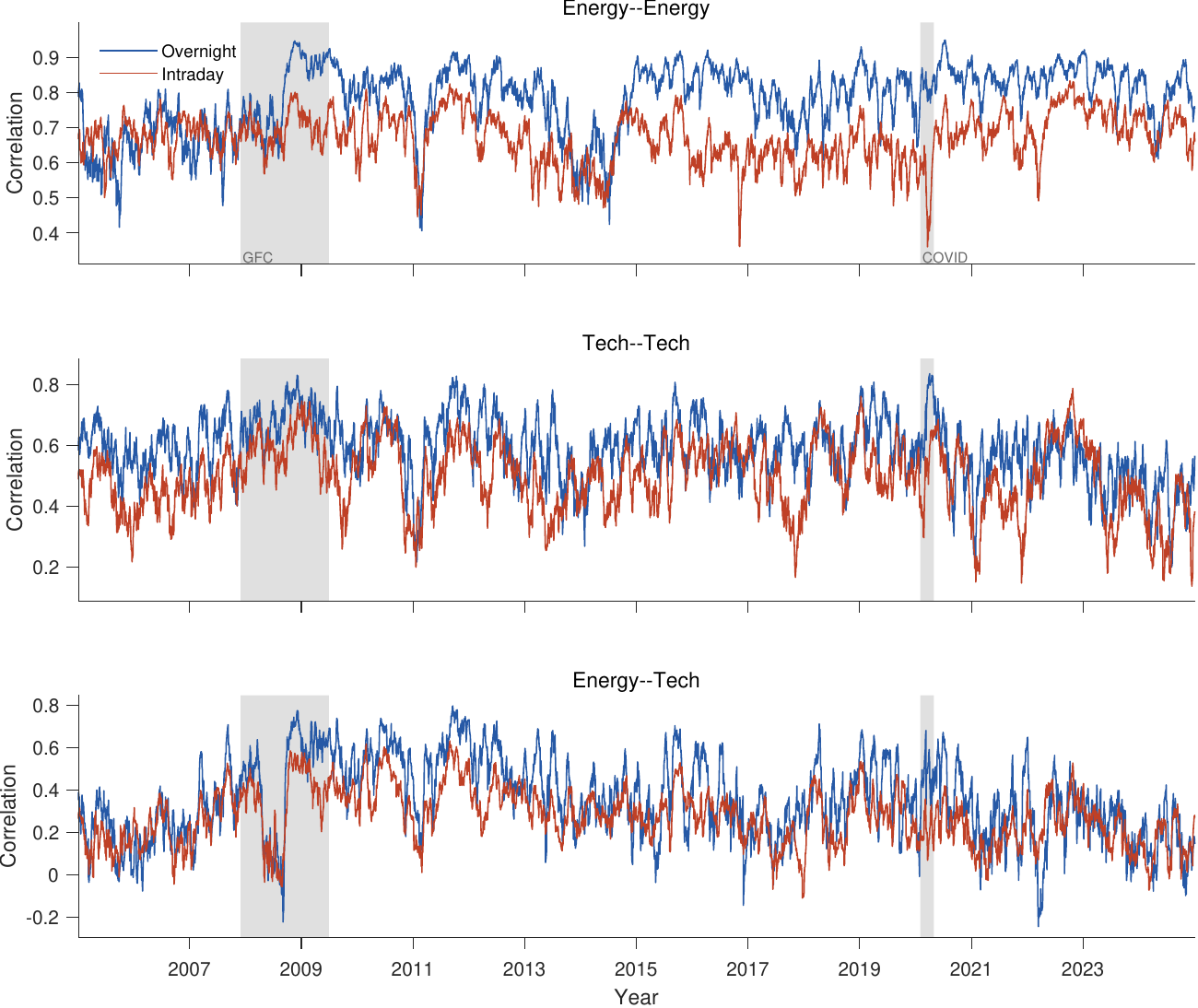}
\vspace{-0.3em}
\caption{Dynamic Block Correlations: Score-Driven Model under Cluster-$t$ Distribution}
\label{fig:Figure2}
\vspace{-0.3em}
\begin{minipage}{0.95\textwidth}
\footnotesize
\textit{Note:} This figure plots the estimated conditional block correlations from the
Score-Block-Cluster-$t$ model with session-by-sector tail clustering $(G=4)$. The upper, middle, and lower panels
report Energy, Technology, and cross-sector correlations, respectively.
Shaded areas indicate NBER-dated recessions.
\end{minipage}
\vspace{0.3em}
\end{figure}

Overall, these results reinforce a consistent message: imposing a simple
sector-based block structure captures most of the relevant dependence in the
data, substantially reducing model complexity while preserving most of the fit,
and often improving information-criterion performance.

To visualize the dependence dynamics implied by the preferred score-driven
block specification, Figure~\ref{fig:Figure2} plots the
estimated conditional block correlations for overnight and intraday returns.
Within the Energy sector, shown in the upper panel, overnight correlations are
consistently higher than their intraday counterparts throughout the sample
period, with the gap remaining stable across different market regimes. Both
series rise sharply during the Great Recession, reflecting increased
co-movement among energy stocks during periods of broad market stress, and
briefly spike at the onset of the COVID-19 shock before returning toward
pre-pandemic levels.

Within the Information Technology sector, shown in the middle panel, overnight
conditional correlations exceed intraday conditional correlations for most of
the sample. The two series also display pronounced time variation, with
correlation increases around major market stress episodes. This pattern is
consistent with the importance of after-hours information arrivals for
technology firms, including earnings announcements and guidance releases, which
can generate substantial overnight dependence dynamics.

Cross-sector correlations, shown in the lower panel, exhibit no systematic gap
between overnight and intraday series, with the two lines frequently crossing
over the sample. During the Great Recession, correlations increase markedly and
become more volatile in both sessions. Outside crisis periods, cross-sector
correlations remain moderate and time-varying, suggesting that sectoral
segmentation captures an important component of the dependence structure while
still allowing for changing common exposure across sectors.

\subsection{Out-of-Sample Evaluation\label{subsec:Six-Asset-OOS}}

For the six-asset sample, we estimate models using data from 2005 to 2021 and
evaluate out-of-sample performance over the period from 2022 to 2024
($T_{oos}=753$). Table \ref{tab:Out-of-Sample-Comparison} reports the
out-of-sample log-likelihoods of total return $\ell$, overnight return
$\ell_{N}$, and intraday return $\ell_{D}$, along with $p$-values from the
Model Confidence Set (MCS) test of \citet{Hansen2011}. Models with an MCS
$p$-value above 0.05 belong to the superior set of models (SSM).

\begin{table}[!t]
\centering
\caption{Out-of-Sample Performance 
\label{tab:Out-of-Sample-Comparison}}
\vspace{0.2em}
\begin{threeparttable}
	\footnotesize
	\setlength{\tabcolsep}{3pt}
	\renewcommand{\arraystretch}{0.86}
	
	\begin{tabularx}{\textwidth}{l *{5}{>{\centering\arraybackslash}X}}
		\toprule
		\midrule
		& \multicolumn{3}{c}{\textbf{Log-Likelihood Components}} 
		& \textbf{Complexity} & \textbf{Stat. Test} \\
		\cmidrule(lr){2-4} \cmidrule(lr){5-5} \cmidrule(lr){6-6}
		Model & $\ell_N$ & $\ell_D$ & $\ell$ & $p$ & MCS $p$-value \\
		\midrule
	
		\multicolumn{6}{l}{\textbf{Panel A: Gaussian Distribution}} \\
		DCC        & -5694 & -5577 & -11271 & 114 & 0.000 \\
		Score-Full & -5708 & -5575 & -11283 & 90  & 0.000 \\
		
		\addlinespace[0.3ex]
		\midrule
		\multicolumn{6}{l}{\textbf{Panel B: Multivariate-$t$ Distribution ($G=1$)}} \\
		DCC         & -4766 & -5637 & -10404 & 115 & 0.000 \\
		Score-Full  & -4713 & -5625 & -10338 & 91  & 0.000 \\
		Score-Block & -4688 & -5624 & -10312 & 19  & 0.000 \\
		
		\addlinespace[0.3ex]
		\midrule
		\multicolumn{6}{l}{\textbf{Panel C: Cluster-$t$ Distribution (Session, $G=2$)}} \\
		DCC         & -4678 & -5475 & -10153 & 116 & 0.000 \\
		Score-Full  & -4629 & -5470 & -10099 & 92  & 0.000 \\
		Score-Block & -4613 & -5486 & -10099 & 20  & 0.000 \\
		
		\addlinespace[0.3ex]
		\midrule
		\multicolumn{6}{l}{\textbf{Panel D: Cluster-$t$ Distribution (Session $\times$ Sector, $G=4$)}} \\
		Score-Full  & -4485 & -5440 & \textbf{-9925} & 94 & \textbf{1.000} \\
		Score-Block & -4478 & -5458 & -9936 & 22 & \textbf{0.498} \\
		
		\addlinespace[0.3ex]
		\midrule
		\multicolumn{6}{l}{\textbf{Panel E: Hetero-$t$ Distribution (Session $\times$ Asset, $G=12$)}} \\
		Score-Full  & -4577 & -5450 & -10027 & 102 & 0.002 \\
		Score-Block & -4555 & -5469 & -10024 & 30  & 0.002 \\
		
		\midrule
		\bottomrule
	\end{tabularx}
	
	\begin{tablenotes}[flushleft]
		\scriptsize
		\setlength{\itemsep}{0pt}
		\item \textit{Note}: This table reports out-of-sample log-likelihoods for the forecasting period 2022--2024 ($T_{\mathrm{oos}}=753$). Models are grouped by distributional assumptions, from Gaussian to Hetero-$t$. $\ell_N$ and $\ell_D$ are decomposed overnight and intraday log-likelihoods, respectively, and $p$ denotes the number of parameters. 
        The Model Confidence Set (MCS) test evaluates predictive ability based on out-of-sample log-likelihoods.
         Bold MCS values indicate models in the superior set (MCS $p$-value $>0.05$); the globally optimal total log-likelihood is also bolded.
	\end{tablenotes}
\end{threeparttable}
\vspace{-0.6em}
\end{table}

The out-of-sample results are broadly consistent with the in-sample findings.
Score-driven specifications tend to outperform the DCC benchmark under
heavy-tailed distributions, and all DCC models are excluded from the SSM. Under
the Gaussian specification, both DCC and score-driven models perform poorly.
Moving from the multivariate-$t$ to the session-level Cluster-$t$ leads to
noticeable improvements in log-likelihood, indicating that allowing for tail
heterogeneity across return components remains important for predictive
performance.

The session-by-sector Cluster-$t$ specifications in Panel D achieve the
strongest out-of-sample performance. Score-Full-Cluster-$t$ attains the highest
total log-likelihood of $-9,925$ and an MCS $p$-value of $1.000$.
Score-Block-Cluster-$t$ also enters the SSM with a $p$-value of $0.498$ and
achieves a competitive log-likelihood of $-9,936$ with only $22$ second-stage
parameters, reflecting the parsimony gains from the sector-based block
structure. These two specifications are the only models not excluded by the MCS
test at the 5\% level.

The Hetero-$t$ specifications in Panel E perform worse out-of-sample than the
session-by-sector Cluster-$t$, with both Score-Full and Score-Block
specifications excluded from the SSM. This suggests that, with only three assets
per sector-session block, the gain from additional marginal tail flexibility
does not compensate for the loss of a shared sector-level tail component. The
result is consistent with the in-sample evidence in Section
\ref{subsec:Six-Asset-Multivariate}: preserving sector-level tail structure is
more valuable than assigning a separate tail parameter to each asset.

The six-asset application provides a detailed setting in which to examine
session-level and sector-level tail heterogeneity, unrestricted versus block
correlation dynamics, and predictive performance. We next turn to a larger
cross section to assess whether the proposed block representation remains
tractable in high-dimensional settings.

\section{High-Dimensional Empirical Application with 100 Assets\label{sec:High-Dimensional-Application}}

The preceding section used a small cross section to examine the mechanisms of
the proposed model in detail. This section studies a 100-asset application to
evaluate scalability. The purpose is not to replace the six-asset analysis, but
to show that the same framework remains computationally tractable and
empirically informative in a substantially larger cross section.

\subsection{Sample Construction\label{subsec:High-Dimensional-Sample}}

The high-dimensional sample consists of 100 U.S. equities grouped into ten
sectors over the period 2005--2023. The earlier
endpoint relative to the six-asset sample reflects the need to maintain a
balanced panel with complete open and close price records for all selected
firms. The full list
of assets and their sector classifications is provided in Appendix
Table~\ref{tab:Stock-List-100}. Since unrestricted correlation models are
computationally impractical in this dimension, the analysis focuses on
score-driven block specifications under alternative tail partitions. While the diagnostic evidence for the block-diagonal cross-session restriction
$C_t^{ND}=0$ is based on the six-asset sample, the same restriction is
maintained in the 100-asset application as a parsimonious parametric assumption
for scalable estimation.

\subsection{In-Sample Block Correlation Estimates\label{subsec:High-Dimensional-InSample}}

Table \ref{tab:Block-Estimation-100} reports the implied block correlations,
dynamic parameters, degrees of freedom, and in-sample fit measures for the
100-asset sample under the score-driven block specifications.

\begin{table}[!t]
\centering
\caption{High-Dimensional Block Correlation Estimates for the 100-Asset Sample
\label{tab:Block-Estimation-100}}
\vspace{0.2em}
\begin{threeparttable}
	\footnotesize
	\setlength{\tabcolsep}{2.2pt}
	\renewcommand{\arraystretch}{0.82}
	
	\begin{tabularx}{\textwidth}{l c YYYYYYYY}
		\toprule
		\midrule
		& & \multicolumn{2}{c}{Score-Block-Multivariate-$t$}
		& \multicolumn{2}{c}{Score-Block-Cluster-$t$}
		& \multicolumn{2}{c}{Score-Block-Cluster-$t$}
		& \multicolumn{2}{c}{Score-Block-Hetero-$t$} \\
		
		& & \multicolumn{2}{c}{}
		& \multicolumn{2}{c}{Session}
		& \multicolumn{2}{c}{Session $\times$ Sector}
		& \multicolumn{2}{c}{Session $\times$ Asset} \\
		
		& & \multicolumn{2}{c}{($G=1$)}
		& \multicolumn{2}{c}{($G=2$)}
		& \multicolumn{2}{c}{($G=20$)}
		& \multicolumn{2}{c}{($G=200$)} \\
		
		\cmidrule(lr){3-4} \cmidrule(lr){5-6} \cmidrule(lr){7-8} \cmidrule(lr){9-10}
		& & Night & Day & Night & Day & Night & Day & Night & Day \\
		\midrule
		
		\multicolumn{10}{l}{\textbf{Panel A: Implied Correlations and Parameter Estimates}} \\
		\midrule
		
\multirow{3}{*}{$\rho$}
& Q25 & 0.4459 & 0.2558 & 0.3593 & 0.3025 & 0.3908 & 0.3176 & 0.3617 & 0.3126 \\
& Q50 & 0.4494 & 0.2581 & 0.3621 & 0.3052 & 0.3955 & 0.3204 & 0.3665 & 0.3158 \\
& Max & 0.4589 & 0.2649 & 0.3719 & 0.3122 & 0.4066 & 0.3282 & 0.3807 & 0.3243 \\
\addlinespace[0.2ex]

\multirow{3}{*}{$\mu$}
& Q25 & 0.0385 & 0.0292 & 0.0364 & 0.0278 & 0.0371 & 0.0327 & 0.0350 & 0.0323 \\
& Q50 & 0.0447 & 0.0364 & 0.0412 & 0.0390 & 0.0423 & 0.0399 & 0.0400 & 0.0399 \\
& Max & 0.0541 & 0.0522 & 0.0508 & 0.0534 & 0.0630 & 0.0574 & 0.0492 & 0.0549 \\
\addlinespace[0.2ex]

\multirow{3}{*}{$\beta$}
& Q25 & 0.7916 & 0.9573 & 0.4052 & 0.8940 & 0.8955 & 0.8107 & 0.9218 & 0.7902 \\
& Q50 & 0.8807 & 0.9835 & 0.8513 & 0.9827 & 0.9705 & 0.9303 & 0.9746 & 0.9402 \\
& Max & 1.0000 & 0.9975 & 1.0000 & 0.9999 & 0.9999 & 0.9976 & 0.9991 & 0.9997 \\
\addlinespace[0.2ex]

\multirow{3}{*}{$\alpha$}
& Q25 & 0.0050 & 0.0044 & 0.0016 & 0.0041 & 0.0085 & 0.0102 & 0.0088 & 0.0084 \\
& Q50 & 0.0143 & 0.0072 & 0.0056 & 0.0088 & 0.0239 & 0.0222 & 0.0150 & 0.0219 \\
& Max & 0.0714 & 0.0259 & 0.0432 & 0.0494 & 0.0678 & 0.0723 & 0.0777 & 0.0831 \\
		
\addlinespace[0.3ex]
\midrule
\multicolumn{10}{l}{\textbf{Panel B: Degrees of Freedom}} \\
\midrule
$\nu_0$ && \multicolumn{2}{c}{8.7419} & 4.4947 & 14.3034 &  &  &  &  \\
$\nu_1$ && \multicolumn{2}{c}{} &  &  & 3.3976 & 11.4409 & 2.4095\tnote{$\dagger$} & 5.9396\tnote{$\dagger$} \\
$\nu_2$ && \multicolumn{2}{c}{} &  &  & 3.3357 & 9.8658  & 2.5848\tnote{$\dagger$} & 6.5395\tnote{$\dagger$} \\
$\nu_3$ && \multicolumn{2}{c}{} &  &  & 2.9988 & 10.5407 & 3.1785\tnote{$\dagger$} & 8.4010\tnote{$\dagger$} \\
$\nu_4$ && \multicolumn{2}{c}{} &  &  & 3.4748 & 10.0935 & 2.6145\tnote{$\dagger$} & 5.1442\tnote{$\dagger$} \\
$\nu_5$ && \multicolumn{2}{c}{} &  &  & 3.4525 & 12.2934 & 2.4448\tnote{$\dagger$} & 6.0608\tnote{$\dagger$} \\
$\nu_6$ && \multicolumn{2}{c}{} &  &  & 3.2292 & 10.3155 & 2.9676\tnote{$\dagger$} & 7.3385\tnote{$\dagger$} \\
$\nu_7$ && \multicolumn{2}{c}{} &  &  & 2.7113 & 7.7214  & 2.5563\tnote{$\dagger$} & 5.5523\tnote{$\dagger$} \\
$\nu_8$ && \multicolumn{2}{c}{} &  &  & 3.3386 & 11.3982 & 2.4315\tnote{$\dagger$} & 4.9518\tnote{$\dagger$} \\
$\nu_9$ && \multicolumn{2}{c}{} &  &  & 3.1948 & 10.8006 & 2.6636\tnote{$\dagger$} & 6.2696\tnote{$\dagger$} \\
$\nu_{10}$ && \multicolumn{2}{c}{} &  &  & 3.1097 & 9.3772 & 2.5279\tnote{$\dagger$} & 6.3601\tnote{$\dagger$} \\
		
\addlinespace[0.3ex]
\midrule
\multicolumn{10}{l}{\textbf{Panel C: Model Fit Diagnostics}} \\
\midrule
$p$ && \multicolumn{2}{c}{331}& \multicolumn{2}{c}{332}& \multicolumn{2}{c}{350}& \multicolumn{2}{c}{530} \\
$\ell$ && \multicolumn{2}{c}{-1124328}& \multicolumn{2}{c}{-1100451}& \multicolumn{2}{c}{-1042293}& \multicolumn{2}{c}{\textbf{-989034}} \\
$\ell_N$ && \multicolumn{2}{c}{-515857}& \multicolumn{2}{c}{-514676}& \multicolumn{2}{c}{-456925}& \multicolumn{2}{c}{\textbf{-408421}} \\
$\ell_D$ && \multicolumn{2}{c}{-608470}& \multicolumn{2}{c}{-585775}& \multicolumn{2}{c}{-585368}& \multicolumn{2}{c}{\textbf{-580613}} \\
BIC && \multicolumn{2}{c}{2251459}& \multicolumn{2}{c}{2203716}& \multicolumn{2}{c}{2087551}& \multicolumn{2}{c}{\textbf{1982558}} \\
		
		\midrule
		\bottomrule
	\end{tabularx}
	
\begin{tablenotes}[flushleft]
	\scriptsize
	\setlength{\itemsep}{0pt}
	\item \textit{Note}: This table reports score-driven model estimates for 100 assets grouped into ten sectors under a sector-based block correlation structure. Panel A summarizes the distribution of implied sector-pair correlations and condensed dynamic parameters across the lower-triangular sector-pair entries. Q25, Q50, and Max denote the 25th percentile, median, and maximum, respectively. $\rho$, $\mu$, $\beta$, and $\alpha$ denote implied unconditional correlations, condensed long-run correlations, persistence parameters, and score sensitivity parameters, respectively. In Panel B, $\nu_0$ denotes the global or session-specific degrees of freedom for models with $G=1$ or $G=2$, while $\nu_1,\ldots,\nu_{10}$ denote sector-specific degrees of freedom under the Session $\times$ Sector Cluster-$t$ model ($G=20$). For the Session $\times$ Asset Hetero-$t$ model ($G=200$), the reported $\nu_1,\ldots,\nu_{10}$\tnote{$\dagger$} are sector-level averages of asset-specific degrees of freedom. $\ell_N$ and $\ell_D$ are decomposed log-likelihoods. The parameter count $p$ includes only second-stage dependence and tail
parameters used in the likelihood comparison and BIC calculation. Bold values indicate the superior specification.

\end{tablenotes}
	
\end{threeparttable}
\vspace{-0.6em}
\end{table}

The results show that the block structure remains computationally tractable in
the larger cross section. Allowing for more flexible tail heterogeneity leads to sizeable
improvements in both log-likelihood and BIC. In particular, the Hetero-$t$
specification delivers the best in-sample fit, with the highest total
log-likelihood and the lowest BIC. This suggests that asset-level tail
heterogeneity becomes more important as the cross-sectional dimension
increases.

\begin{figure}
\centering
\vspace{-0.6em}
\includegraphics[width=0.9\textwidth]{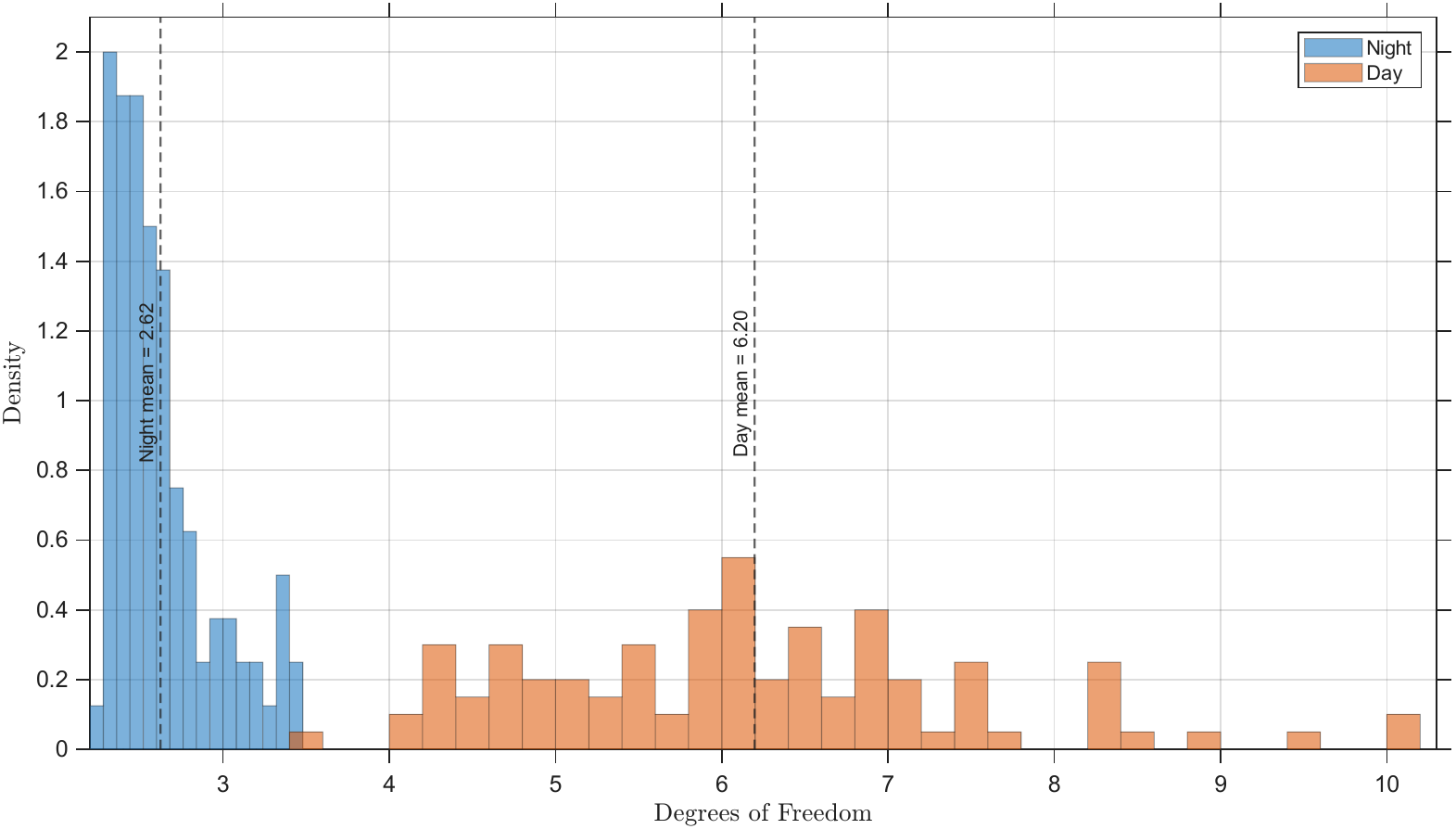}
\vspace{-0.6em}
\caption{Distribution of Asset-Level Degrees of Freedom under the Hetero-$t$ Specification}
\label{fig:HeteroT-DF-100}
\vspace{-0.3em}
\begin{minipage}{0.9\textwidth}
\footnotesize
\textit{Note}: This figure plots the distribution of asset-level degrees of
freedom estimated under the Score-Block-Hetero-$t$ specification in the
100-asset sample. Lower values correspond to heavier tails.
\end{minipage}
\vspace{-0.2em}
\end{figure}

The estimated degrees of freedom reveal a clear session contrast. The
session-level Cluster-$t$ estimates differ in absolute magnitude from those in
the six-asset application. This difference is expected because the 100-asset
sample covers a broader and more diversified set of firms and sectors. This session contrast remains visible under the more flexible specifications. Across the
sector-level and asset-level specifications, the overnight degrees of freedom
are systematically lower than their intraday counterparts, indicating heavier
tails in overnight innovations. Figure~\ref{fig:HeteroT-DF-100} further
illustrates this pattern under the Hetero-$t$ specification. The overnight
degrees of freedom are concentrated at lower values than the intraday degrees
of freedom, providing visual evidence of stronger tail thickness in overnight
innovations. 

Overall, the 100-asset in-sample results confirm that the sector-based block structure provides a parsimonious and scalable representation of
high-dimensional dependence. It substantially reduces the dimensionality of the
correlation dynamics while preserving economically meaningful variation in both
correlations and tail behavior.

\subsection{Out-of-Sample Evaluation\label{subsec:High-Dimensional-OOS}}

We also examine out-of-sample performance in the 100-asset sample. Models are estimated using data from 2005--2020 and evaluated over 2021--2023. Since unrestricted correlation models are computationally impractical in this high-dimensional
setting, the comparison focuses on score-driven block models under alternative
tail specifications. Table \ref{tab:OOS-Block-100} reports both the likelihood-based evaluation and
the correlation-based GMV portfolio evaluation.

For the portfolio evaluation, we construct correlation-based global minimum
variance (GMV) portfolios using the model-implied forecasts of the joint
split-session correlation matrix. Let $C_{m,t+1|t}$ denote the one-step-ahead
conditional correlation matrix forecast from model $m$, where
$C_{m,t+1|t}=\mathrm{blockdiag}(C^N_{m,t+1|t},C^D_{m,t+1|t})$. Since the
second-stage model is specified for standardized innovations, this exercise uses
the forecast correlation matrix rather than the full return covariance matrix.
The corresponding GMV weights are
\[
\widehat{w}_{m,t+1|t}
=
\frac{C_{m,t+1|t}^{-1}\mathbf{1}}
{\mathbf{1}^{\prime}C_{m,t+1|t}^{-1}\mathbf{1}} .
\]
The realized portfolio return is computed as
$\widehat{w}_{m,t+1|t}^{\prime}Z_{t+1}$, where
$Z_{t+1}=(Z_{t+1}^{N\prime},Z_{t+1}^{D\prime})^{\prime}$ denotes the vector of
standardized overnight and intraday innovations. We compare models using the
realized variance of these portfolio returns and apply the MCS test to the
corresponding squared portfolio returns. This exercise is therefore interpreted
as a correlation-based evaluation of dependence forecasts rather than a full
volatility-scaled asset-allocation exercise.

The high-dimensional out-of-sample results broadly support the in-sample
evidence. More flexible tail specifications improve predictive performance
relative to the common multivariate-$t$ benchmark, and the Hetero-$t$ model
delivers the highest total log-likelihood. It is also the only model retained in the superior set by the likelihood-based
MCS test. The GMV results are consistent with this ranking: the Hetero-$t$
model delivers the lowest realized variance and is retained in the superior set
by the GMV MCS test. This out-of-sample evidence suggests that
the additional flexibility of the Hetero-$t$ specification captures persistent
asset-level tail heterogeneity rather than merely fitting in-sample noise.  In contrast to the six-asset benchmark,
where the session-by-sector Cluster-$t$ performs best, the 100-asset results
suggest that asset-level tail heterogeneity becomes more valuable as the
cross-sectional dimension increases.


\begin{table}[H]
\centering
\caption{High-Dimensional Out-of-Sample Performance and GMV Evaluation
\label{tab:OOS-Block-100}}
\vspace{0.2em}
\begin{threeparttable}
	\footnotesize
	\setlength{\tabcolsep}{3.5pt}
	\renewcommand{\arraystretch}{0.88}
	
	\begin{tabularx}{\textwidth}{l *{7}{>{\centering\arraybackslash}X}}
		\toprule
		\midrule
		& \multicolumn{5}{c}{\textbf{Out-of-Sample Log-Likelihood Evaluation}} 
		& \multicolumn{2}{c}{\textbf{GMV Performance}} \\
		\cmidrule(lr){2-6} \cmidrule(lr){7-8}
		Model & $\ell_N$ & $\ell_D$ & $\ell$ & $p$ & MCS-$p$
		& RV & MCS-$p$ \\
		\midrule
		
		\multicolumn{8}{l}{\textbf{Panel A: Score-Block-Multivariate-$t$ Distribution ($G=1$)}} \\
		Score-Block 
		& -86629 
		& -98042 
		& -184671 
		& 331 
		& 0.000
		& 0.1519
		& 0.013 \\
		
		\addlinespace[0.3ex]
		\midrule
		\multicolumn{8}{l}{\textbf{Panel B: Score-Block-Cluster-$t$ Distribution (Session, $G=2$)}} \\
		Score-Block 
		& -85938 
		& -94512 
		& -180450 
		& 332 
		& 0.000
		& 0.1429
		& 0.183 \\
		
		\addlinespace[0.3ex]
		\midrule
		\multicolumn{8}{l}{\textbf{Panel C: Score-Block-Cluster-$t$ Distribution (Session $\times$ Sector, $G=20$)}} \\
		Score-Block 
		& -78624 
		& -94627 
		& -173250 
		& 350 
		& 0.000
		& 0.1402
		& 0.183 \\
		
		\addlinespace[0.3ex]
		\midrule
		\multicolumn{8}{l}{\textbf{Panel D: Score-Block-Hetero-$t$ Distribution (Session $\times$ Asset, $G=200$)}} \\
		Score-Block 
		& \textbf{-72665} 
		& \textbf{-93679} 
		& \textbf{-166344} 
		& 530 
		& \textbf{1.000}
		& \textbf{0.1376}
		& \textbf{1.000} \\
		
		\midrule
		\bottomrule
	\end{tabularx}
	
	\begin{tablenotes}[flushleft]
		\scriptsize
		\setlength{\itemsep}{0pt}
		\item \textit{Note}: This table reports out-of-sample log-likelihoods and GMV portfolio performance for the forecasting period 2021--2023. The models are estimated under a sector-based block correlation structure for 100 assets grouped into ten sectors. $\ell_N$ and $\ell_D$ denote the overnight and intraday log-likelihood components, respectively, and $\ell$ denotes the total out-of-sample log-likelihood. For the Score-Block-Multivariate-$t$ model, $\ell$ is computed from the joint likelihood and therefore does not necessarily equal the sum of the reported decomposed components. The number of parameters is denoted by $p$. The log-likelihood MCS test is based on out-of-sample log-likelihood losses. The GMV realized variance is computed from the model-implied joint split-session correlation matrix, $C_{t}=\mathrm{blockdiag}(C^N_t,C^D_t)$, and the standardized innovation vector
$Z_t=(Z_t^{N\prime},Z_t^{D\prime})^{\prime}$. The GMV MCS test is based on squared portfolio returns. Bold values indicate the highest log-likelihood, the lowest GMV realized variance, and models included in the superior set according to the corresponding MCS test.
	\end{tablenotes}
\end{threeparttable}
\vspace{-0.6em}
\end{table}

\section{Conclusion\label{sec:Conclusion}}

This paper develops Split-Session Cluster GARCH, a model for heavy-tailed
multivariate dependence in asset returns decomposed into overnight and intraday
components. The model combines asset-specific Coupled EGARCH dynamics with a
score-driven correlation model for standardized innovations. Tail heterogeneity
is introduced through convolution-$t$ distributions, allowing degrees of freedom
to vary across clusters defined by trading sessions and economically meaningful
asset groups. The correlation dynamics use the unconstrained correlation
parameterization and the canonical block-correlation representation, preserving
positive definiteness while improving scalability.

The empirical results show that separating overnight and intraday returns is
important for modeling heavy-tailed dependence. In the six-asset application,
overnight innovations are substantially more heavy-tailed than intraday
innovations, with estimated degrees of freedom differing by roughly a factor of
three. A common multivariate-$t$ specification masks this session-level tail
heterogeneity. Allowing separate tail parameters by session improves fit, and
further partitioning tails by sector delivers additional gains concentrated
mainly in the overnight component. These findings indicate that tail behavior
varies along both time and cross-sectional dimensions.

The results also show that score-driven updating is most effective when paired
with an appropriate heavy-tailed specification. Relative to traditional DCC
models, score-driven specifications perform better under heavy-tailed
distributions, especially for overnight returns, where extremes are more
frequent. The clustered convolution-$t$ specification localizes the impact of
extremes through session- and group-specific tail weights.

The 100-asset application confirms that the framework remains feasible in a
larger cross section. The block representation reduces the dependence system
from the asset-pair level to the block-pair level while still allowing rich tail
heterogeneity. In this setting, the block Hetero-$t$ specification delivers the strongest
in-sample and out-of-sample likelihood performance among the block models and
also achieves the lowest realized variance in the GMV portfolio evaluation.

Overall, imposing a common tail structure can distort dynamic correlation
modeling when tail behavior differs across sessions and asset groups.
Split-Session Cluster GARCH addresses this issue by combining session-specific
volatility dynamics, clustered heavy-tailed innovations, and scalable
block-structured correlation dynamics. Future research could consider
data-driven clustering, time-varying cluster memberships, broader asset
universes, and high-dimensional asymptotic theory.

\FloatBarrier

\appendix
\numberwithin{equation}{section}
\numberwithin{table}{section}
\numberwithin{figure}{section}

\renewcommand{\theequation}{\Alph{section}.\arabic{equation}}
\renewcommand{\thetable}{\Alph{section}.\arabic{table}}
\renewcommand{\thefigure}{\Alph{section}.\arabic{figure}}

\newpage

\section*{Appendix}
\section{Composition of the 100-Asset Sample}
\label{sec:Composition-100-Asset-Sample}

Table~\ref{tab:Stock-List-100} lists the 100 stocks used in the
high-dimensional analysis. The stocks are grouped by sector according to their
sector classifications in the sample.
\vspace{0.4em}

\begingroup
\refstepcounter{table}
\label{tab:Stock-List-100}
\noindent\textbf{Table~\thetable}\\
\noindent List of Stocks in the 100-Asset Sample
\begin{center}
\begin{threeparttable}
\footnotesize
\setlength{\tabcolsep}{4pt}
\renewcommand{\arraystretch}{1.05}

\begin{tabularx}{\textwidth}{p{0.18\textwidth} X  p{0.15\textwidth} X}
\toprule
\midrule
\textbf{Sector} & \textbf{Stocks} & \textbf{Sector} & \textbf{Stocks} \\
\midrule

Information Technology 
& AAPL, ACN, ADBE, CRM, \textbf{CSCO}, IBM, \textbf{INTC}, \textbf{MSFT}, NVDA, ORCL, QCOM, TXN, XRX
& Energy 
& \textbf{APA}, BKR, COP, \textbf{CVX}, \textbf{DVN}, HAL, MRO, NOV, OXY, SLB, WMB, XOM \\

\addlinespace[0.35ex]

Healthcare 
& ABT, AMGN, BAX, BMY, DHR, GILD, JNJ, LLY, MDT, MRK, PFE, TMO, UNH
& Materials 
& APD, ECL, FCX, IP, SHW \\

\addlinespace[0.35ex]

Utilities 
& AEE, AEP, D, DUK, ETR, EXC, SO
& Industrials 
& BA, CAT, EMR, FDX, GD, GE, HON, LMT, MMM, NSC, UNP, UPS \\

\addlinespace[0.35ex]

Financials 
& ALL, AXP, BAC, BK, C, COF, GS, JPM, MET, RF, USB, WFC
& Consumer Staples 
& CL, COST, CPB, KO, MDLZ, MO, PEP, PG, WBA, WMT \\

\addlinespace[0.35ex]

Consumer Discretionary 
& AMZN, EBAY, F, HD, LOW, MCD, NKE, SBUX, TGT
& Telecom. Services 
& CMCSA, DIS, DISH, NFLX, OMC, T, VZ \\
\midrule
\bottomrule
\end{tabularx}

\begin{tablenotes}[flushleft]
\scriptsize
\setlength{\itemsep}{0pt}
\item \textit{Note}: This table reports the composition of the 100-asset sample by sector. Stocks are grouped according to their Global Industry Classification Standard (GICS) sector classifications. Bold entries indicate the six stocks used in the small sample.
\end{tablenotes}

\end{threeparttable}

\end{center}
\endgroup

\section{Diagnostic Test}
\label{sec:Diagnostic-Test-on}

Table \ref{tab:Contemporaneous-Correlation} reports diagnostic checks of
residual contemporaneous cross-session dependence between the standardized
overnight and intraday innovations. Since the Coupled EGARCH framework
explicitly captures the most direct same-asset cross-session dependence
through the transmission parameter $\delta_i$ and the coupled variance
dynamics, the standardized residuals are expected to exhibit little remaining
linear contemporaneous dependence across sessions.

\begingroup
\refstepcounter{table}
\label{tab:Contemporaneous-Correlation}
\vspace{0.5em}
\noindent\textbf{Table~\thetable}\\
\noindent Contemporaneous Correlation
\begin{center}
\begin{threeparttable}
	\footnotesize
	\setlength{\tabcolsep}{4pt}
	\renewcommand{\arraystretch}{1.05}

	\begin{tabularx}{\textwidth}{lYYY}
		\toprule
		\toprule
		\multicolumn{4}{l}{\textbf{Panel A: Same-Asset Cross-Session Correlations}} \\
		\midrule
		Asset & $\rho_{ND}$ & $t$-stat (Parzen) & $p$-value \\
		\midrule
		\multicolumn{4}{l}{\textit{Energy Sector}} \\
		CVX  & 0.0060 & 0.411 & 0.681 \\
		APA  & 0.0218 & 1.118 & 0.263 \\
		DVN  & 0.0188 & 1.208 & 0.227 \\
		\addlinespace[0.4ex]
		\multicolumn{4}{l}{\textit{Technology Sector}} \\
		MSFT & 0.0067 & 0.419 & 0.675 \\
		INTC & 0.0299 & 1.789 & 0.074 \\
		CSCO & 0.0140 & 0.811 & 0.417 \\

		\addlinespace[1.0ex]
		\midrule
		\multicolumn{4}{l}{\textbf{Panel B: Cross-Asset Cross-Session Correlation Ranges}} \\
		\midrule
		Night sector / Day sector 
		& Correlation range 
		& $t$-stat range (Parzen) 
		& $p$-value range \\
		\midrule
		Energy / Energy 
		& [0.0140,0.0656]   
		& [0.989,4.195]    
		& [0.000,0.323] \\

		Energy / Technology 
		& [-0.0083,0.0238]  
		& [-0.564,1.391]   
		& [0.164,0.986] \\

		Technology / Energy 
		& [0.0075,0.0380]   
		& [0.517,2.561]    
		& [0.010,0.605] \\

		Technology / Technology 
		& [-0.0047,0.0619] 
		& [-0.252,4.089]   
		& [0.000,0.801] \\
		\midrule
		\bottomrule
	\end{tabularx}

	\begin{tablenotes}[flushleft]
		\footnotesize
	\item \textit{Note}: Panel A reports same-asset Pearson correlations,
$\rho(Z_{i,t}^{N},Z_{i,t}^{D})$. Panel B summarizes cross-asset cross-session
correlations, $\rho(Z_{i,t}^{N},Z_{j,t}^{D})$ for $i\neq j$, grouped by the
overnight and intraday asset sectors. Same-asset pairs are excluded from
Panel B. HAC standard errors with the Parzen kernel and bandwidth $M=5$ are
used to compute the associated $t$-statistics and $p$-values. Values reported
as $0.000$ are smaller than $0.0005$.
	\end{tablenotes}
\end{threeparttable}
\end{center}
\vspace{0.3em}
\endgroup

Consistent with this expectation, Panel A shows that the same-asset
overnight-intraday correlations are uniformly small, with absolute values
below 0.03. Using HAC standard errors with the Parzen kernel and bandwidth
$M=5$, none of these correlations is
statistically significant at the $5\%$ level. The largest same-asset estimate
is observed for INTC, with a correlation of 0.0299, which is economically
negligible.

Panel B further summarizes cross-asset cross-session correlations of the form
$\rho(Z_{i,t}^{N},Z_{j,t}^{D})$ for $i\neq j$, grouped by the sector of the
overnight asset and the sector of the intraday asset. These correlations also
remain small in magnitude. The within-sector cross-session ranges are
[0.0140,0.0656] for Energy and [-0.0047,0.0619] for Technology, while the
cross-sector ranges are [-0.0083,0.0238] for Energy overnight-Technology
intraday pairs and [0.0075,0.0380] for Technology overnight-Energy intraday
pairs. Based on Parzen-kernel HAC inference, across all 30 cross-asset cross-session pairs, 21 are statistically
insignificant at the $5\%$ level, and 27 have absolute correlations below
$0.05$. Even the largest cross-asset cross-session correlation, $0.0656$, is
much smaller than the within-session correlations in
Table~\ref{tab:Session-Correlation}.

Overall, the evidence suggests that the remaining linear contemporaneous
dependence between overnight and intraday standardized residuals is
economically limited, which is consistent with the block-diagonal correlation
specification adopted in the correlation model. These diagnostics are based on
linear correlations and therefore should be interpreted as evidence on residual
linear cross-session dependence; they do not test for nonlinear cross-session
dependence or tail co-movement.

\section{The Score and Fisher Information for the General Specification \label{sec:The-Score-and}}

In this appendix, we provide the log-likelihood, the score vector,
and the Fisher information matrix for the general $2n\times2n$ joint
conditional correlation matrix $C_{t}$ under the multivariate-$t$
and convolution-$t$ distributions.

Following \citet{Archakov2021}, we parameterize $C_{t}$ using
$\gamma_t=\operatorname{vecl}(\log C_t)$, where $\operatorname{vecl}(\cdot)$
stacks the strictly lower triangular elements of a matrix into a column
vector. We define the Jacobian matrix as $M_{t}=\partial\operatorname{vec}(C_{t})/\partial\gamma_{t}^{\prime}$.
For brevity in the derivation, we omit the time subscript $t$ in
the following expressions.

\subsection{The General Multivariate-$t$ Distribution}

Assume the $2n\times1$ joint innovation vector $Z$ follows a standardized
multivariate-$t$ distribution with $\nu$ degrees-of-freedom, $Z\sim t_{\nu}^{\text{std}}\left(\mathbf{0},C\right)$.
The conditional log-likelihood function is:
\begin{equation}
\ell\left(Z\right)=c_{\nu,2n}-\frac{1}{2}\log|C|-\frac{\nu+2n}{2}\log\left(1+\frac{1}{\nu-2}Z^{\prime}C^{-1}Z\right),
\end{equation}

\noindent where $c_{\nu,2n}$ is the normalizing constant. Define $W=\frac{\nu+2n}{\nu-2+Z^{\prime}C^{-1}Z}$. The
score vector with respect to $\gamma$ is
\begin{equation}
\nabla=\frac{\partial\ell}{\partial\gamma}=\frac{1}{2}M^{\prime}\left(C^{-1}\otimes C^{-1}\right)\left[W\operatorname{vec}\left(ZZ^{\prime}\right)-\operatorname{vec}\left(C\right)\right].
\end{equation}

The corresponding conditional Fisher information matrix $\mathcal{I}=\mathbb{E}\left[\nabla\nabla^{\prime}\right]$
is
\begin{equation}
\mathcal{I}=\frac{1}{4}M^{\prime}\left[\phi\left(C^{-1}\otimes C^{-1}\right)H_{2n}+(\phi-1)\operatorname{vec}\left(C^{-1}\right)\operatorname{vec}\left(C^{-1}\right)^{\prime}\right]M,
\end{equation}

\noindent where $\phi=\frac{\nu+2n}{\nu+2n+2}$, and $H_{2n}=I_{4n^{2}}+K_{2n}$,
where $K_{2n}$ is the $4n^{2}\times4n^{2}$
commutation matrix.

\subsection{The General Convolution-$t$ Distribution}

We consider the more flexible Convolution-$t$ distribution, which
accommodates heterogeneous tail behaviors. The joint innovation vector
is represented as $Z=C^{1/2}U$, where $U$ consists of $G$ independent
multivariate $t$-distributed components $U_{g}$ with dimension $m_{g}$ and
degrees-of-freedom $\nu_{g}$. The log-likelihood function is
\[
\ell(Z)
=
-\frac{1}{2}\log|C|
+
\sum_{g=1}^{G}
\left[
c_g
-\frac{\nu_g+m_g}{2}
\log\left(1+\frac{U_g'U_g}{\nu_g-2}\right)
\right],
\]

\noindent where $U_{g}=E_{g}^{\prime}C^{-1/2}Z$ and $E_{g}\in\mathbb{R}^{2n\times m_{g}}$
is defined by $I_{2n}=\left(E_{1}, \ldots,E_{G}\right)$.
Define $W_g=(\nu_g+m_g)/(\nu_g-2+U_g'U_g)$. The score is
\[
\nabla=M'\Omega'\nabla_s,
\qquad
\nabla_s=\sum_{g=1}^G W_g\operatorname{vec}(E_gU_gU')-\operatorname{vec}(I_{2n}),
\]
with $ \Omega=(I_{2n}\otimes C^{-1/2})(C^{1/2}\oplus I_{2n})^{-1} $. The Fisher information matrix is $
\mathcal{I}=M^{\prime} \Omega^\prime \left(K_{2 n}+\Upsilon_G\right) \Omega M$,
where $\Upsilon_G=\sum_{g=1}^G\Psi_g$ and
\[
\Psi_g
=
\psi_g(I_{2n}\otimes J_g)
+(\phi_g-\psi_g)(J_g\otimes J_g)
+(\phi_g-1)\left[(J_g\otimes J_g)K_{2n}
+\operatorname{vec}(J_g)\operatorname{vec}(J_g)'\right].
\]
Here $J_g=E_gE_g'$, $\phi_g=(\nu_g+m_g)/(\nu_g+m_g+2)$, and
$\psi_g=\phi_g\nu_g/(\nu_g-2)$.

\section{Sector-Based Clustering\label{sec:Sector-Based-Clustering}}

As an alternative specification, we examine whether sector classification
provides a better clustering structure than the session-based decomposition
used in the main analysis. This comparison helps distinguish whether
cross-sectional heterogeneity across industries or temporal heterogeneity
across trading sessions is more important for tail behavior.

We estimate a sector-based Cluster-$t$ model in which the orthogonalized innovation vector $U_t = C_t^{-1/2}Z_t\in\mathbb{R}^{2n}$ is partitioned by sectors rather than by
sessions. Each sector forms an independent cluster containing both overnight
and intraday innovations. For our sample of $n=6$ assets, the partition is
\[
U_{1,t}=\left[U_t^{N,\mathrm{Energy}},U_t^{D,\mathrm{Energy}}\right]^\prime \sim
t_{\nu_1}^{\mathrm{std}}(0,I_6),\qquad
U_{2,t}=\left[U_t^{N,\mathrm{Tech}},U_t^{D,\mathrm{Tech}}\right]^\prime \sim
t_{\nu_2}^{\mathrm{std}}(0,I_6),
\]
where $U_t^{N,s}$ and $U_t^{D,s}$ denote overnight and intraday innovations
for sector $s$ from the partition of the orthogonalized innovation vector $U_t$, and $\nu_1,\nu_2$ are sector-specific degrees of freedom.
Thus, shocks within a sector jointly determine the score downweighting,
regardless of whether they occur overnight or intraday.

Table \ref{tab:Session-and-sector} compares this sector-based partition
with the baseline session-based specification. The results clearly favor
the temporal decomposition: the sector-based model delivers a lower
log-likelihood and a higher BIC. By pooling overnight and intraday
innovations within each sector, it estimates intermediate degrees of freedom
($\nu_1=5.7955$ and $\nu_2=4.2819$), averaging severe overnight tails with
milder intraday tails. This deterioration indicates that session-level heterogeneity is the primary
dimension of tail variation. Sectoral heterogeneity remains important, but it is
most informative when modeled within each trading session rather than by pooling
overnight and intraday innovations within sectors.

\begin{table}[!htbp]
\centering
\caption{Session and Sector Comparison}
\label{tab:Session-and-sector}
\vspace{0.3em}
\begin{threeparttable}
	\footnotesize
	\setlength{\tabcolsep}{2.2pt}
	\renewcommand{\arraystretch}{0.78}
	
	\begin{tabularx}{\textwidth}{l c YYYY}
		\toprule
		\midrule
		& & \multicolumn{2}{c}{Score-Cluster-$t$} 
		  & \multicolumn{2}{c}{Score-Cluster-$t$} \\
		& & \multicolumn{2}{c}{(Session, $G=2$)} 
		  & \multicolumn{2}{c}{(Sector, $K=2$)} \\
		\cmidrule(lr){3-4} \cmidrule(lr){5-6}
		& & Night & Day & Sector 1 & Sector 2 \\
		\midrule
		
		\multicolumn{6}{l}{\textbf{Panel A: Parameter Estimates}} \\
			\midrule
		\multirow{3}{*}{$\mu$}
		& $Q_{25}$ & 0.1660 & 0.1284 & 0.0201 & 0.0140 \\
		& $Q_{50}$ & 0.1970 & 0.1736 & 0.1012 & 0.0319 \\
		& Max      & 0.8313 & 0.7923 & 0.9999 & 0.7342 \\
		
		\addlinespace[0.2ex]
		\multirow{3}{*}{$\beta$}
		& $Q_{25}$ & 0.9910 & 0.9764 & 0.9897 & 0.9134 \\
		& $Q_{50}$ & 0.9950 & 0.9860 & 0.9939 & 0.9713 \\
		& Max      & 0.9995 & 0.9975 & 1.0000 & 0.9969 \\
		
		\addlinespace[0.2ex]
		\multirow{3}{*}{$\alpha$}
		& $Q_{25}$ & 0.0065 & 0.0073 & 0.0018 & 0.0000 \\
		& $Q_{50}$ & 0.0111 & 0.0107 & 0.0043 & 0.0030 \\
		& Max      & 0.0608 & 0.0330 & 0.0299 & 0.0657 \\
		
		\addlinespace[0.3ex]
		\midrule
		\multicolumn{6}{l}{\textbf{Panel B: Degrees of Freedom}} \\
		\midrule
		$\nu$ && 3.3305 & 10.9582 & 5.7955 & 4.2819 \\
		
		\addlinespace[0.3ex]
		\midrule
		\multicolumn{6}{l}{\textbf{Panel C: Model Fit Diagnostics}} \\
		\midrule
		$p$    && \multicolumn{2}{c}{92}      & \multicolumn{2}{c}{92} \\
		$\ell$ && \multicolumn{2}{c}{\textbf{-65242}}  & \multicolumn{2}{c}{-67057} \\
		$\ell_N$ / $\ell_1$ && \multicolumn{2}{c}{-28739}  & \multicolumn{2}{c}{-32298} \\
		$\ell_D$ / $\ell_2$ && \multicolumn{2}{c}{-36503}  & \multicolumn{2}{c}{-34758} \\
		BIC     && \multicolumn{2}{c}{\textbf{131268}}  & \multicolumn{2}{c}{134898} \\
		\midrule
		\bottomrule
	\end{tabularx}
	
	\begin{tablenotes}[flushleft]
		\scriptsize
		\setlength{\itemsep}{0pt}
		\item \textit{Note}: This table compares the session-based Cluster-$t$ ($G=2$) and sector-based Cluster-$t$ ($K=2$) specifications within the score-driven framework. $\mu$ denotes the mean parameter in the transformed correlation space, while
        $\beta$ and $\alpha$ denote persistence and score sensitivity parameters,
        respectively. $Q_{25}$ and $Q_{50}$ denote the 25th and 50th percentiles. $\ell_N/\ell_1$ and $\ell_D/\ell_2$ are decomposed log-likelihoods under the corresponding session- or sector-level partitions. The parameter count $p$ includes only second-stage dependence and tail
parameters used in the likelihood comparison and BIC calculation. Bold values indicate the superior specification.
	\end{tablenotes}
\end{threeparttable}
\vspace{-0.6em}
\end{table}

\section*{Declaration of competing interest}
The authors declare that they have no known competing financial interests or personal relationships that could have appeared to influence the work reported in this paper.

\section*{Data availability}
The data are available from CRSP. Empirical code is available from the authors upon reasonable request.

\section*{Declaration of generative AI and AI-assisted technologies}
During the preparation of this work the authors used OpenAI's ChatGPT in order to assist with language editing, LaTeX drafting, notation checks, and final-readiness review. After using these tools, the authors reviewed and edited the content as necessary and take full responsibility for the content of the publication.

\bibliographystyle{plainnat}
\bibliography{correlationref}

@article{Aielli2013,
  title = {Dynamic Conditional Correlation: On Properties and Estimation},
  author = {Aielli, Gian Piero},
  year = 2013,
  journal = {Journal of Business \& Economic Statistics},
  volume = {31},
  number = {3},
  doi = {10.1080/07350015.2013.771027},
  pages = {282--299},
  publisher = {[American Statistical Association, Taylor \& Francis, Ltd.]},
  urldate = {2026-03-12}
}

@article{Archakov2021,
  title = {A New Parametrization of Correlation Matrices},
  author = {Archakov, Ilya and Hansen, Peter Reinhard},
  year = 2021,
  journal = {Econometrica},
  volume = {89},
  number = {4},
  pages = {1699--1715},
  doi = {10.3982/ECTA16910},
  urldate = {2026-03-03},
  langid = {english}
}

@article{Archakov2024,
  title = {A Canonical Representation of Block Matrices with Applications to Covariance and Correlation Matrices},
  author = {Archakov, Ilya and Hansen, Peter Reinhard},
  year = 2024,
  month = jul,
  journal = {Review of Economics and Statistics},
  volume = {106},
  number = {4},
  pages = {1099--1113},
  doi = {10.1162/rest_a_01258},
  urldate = {2026-03-03},
  langid = {english}
}

@article{Archakov2026,
  title = {A Multivariate Realized {{GARCH}} Model},
  author = {Archakov, Ilya and Hansen, Peter Reinhard and Lunde, Asger},
  year = 2026,
  month = mar,
  journal = {Journal of Econometrics},
  volume = {254},
  pages = {106040},
  doi = {10.1016/j.jeconom.2025.106040},
  urldate = {2026-03-03},
  langid = {english}
}

@article{Barclay2003,
  title = {Price Discovery and Trading After Hours},
  author = {Barclay, Michael J. and Hendershott, Terrence},
  year = 2003,
  month = oct,
  journal = {The Review of Financial Studies},
  volume = {16},
  number = {4},
  pages = {1041--1073},
  doi = {10.1093/rfs/hhg030},
  urldate = {2026-03-15}
}

@article{Blanc2014a,
  title = {The Fine Structure of Volatility Feedback {{II}}: {{Overnight}} and Intra-Day Effects},
  shorttitle = {The Fine Structure of Volatility Feedback {{II}}},
  author = {Blanc, Pierre and Chicheportiche, R{\'e}my and Bouchaud, Jean-Philippe},
  year = 2014,
  month = may,
  journal = {Physica A: Statistical Mechanics and its Applications},
  volume = {402},
  pages = {58--75},
  doi = {10.1016/j.physa.2014.01.047},
  urldate = {2026-03-15}
}

@article{Bollerslev1990,
  title = {Modelling the Coherence in Short-Run Nominal Exchange Rates: A Multivariate Generalized {ARCH} Model},
  author = {Bollerslev, Tim},
  year = 1990,
  journal = {The Review of Economics and Statistics},
  volume = {72},
  number = {3},
  pages = {498--505},
  publisher = {The MIT Press},
  doi = {10.2307/2109358},
  urldate = {2026-03-12}
}

@article{Creal2011,
  title = {A {{Dynamic Multivariate Heavy-Tailed Model}} for {{Time-Varying Volatilities}} and {{Correlations}}},
  author = {Creal, Drew and Koopman, Siem Jan and Lucas, Andr{\'e}},
  year = 2011,
  month = oct,
  journal = {Journal of Business \& Economic Statistics},
  volume = {29},
  number = {4},
  pages = {552--563},
  doi = {10.1198/jbes.2011.10070},
  urldate = {2026-03-03},
  langid = {english}
}

@article{Creal2013,
  title = {Generalized Autoregressive Score Models With Applications},
  author = {Creal, Drew and Koopman, Siem Jan and Lucas, Andr{\'e}},
  year = 2013,
  month = aug,
  journal = {Journal of Applied Econometrics},
  volume = {28},
  number = {5},
  pages = {777--795},
  doi = {10.1002/jae.1279},
  urldate = {2026-03-03},
  copyright = {http://onlinelibrary.wiley.com/termsAndConditions\#vor},
  langid = {english}
}

@article{Creal2015,
  title = {High Dimensional Dynamic Stochastic Copula Models},
  author = {Creal, Drew D. and Tsay, Ruey S.},
  year = 2015,
  month = dec,
  journal = {Journal of Econometrics},
  series = {Frontiers in {{Time Series}} and {{Financial Econometrics}}},
  volume = {189},
  number = {2},
  pages = {335--345},
  doi = {10.1016/j.jeconom.2015.03.027},
  urldate = {2026-03-15}
}

@article{Dhaene2020,
  title = {Incorporating Overnight and Intraday Returns into Multivariate {{GARCH}} Volatility Models},
  author = {Dhaene, Geert and Wu, Jianbin},
  year = 2020,
  month = aug,
  journal = {Journal of Econometrics},
  volume = {217},
  number = {2},
  pages = {471--495},
  doi = {10.1016/j.jeconom.2019.12.013},
  urldate = {2026-03-03},
  langid = {english}
}

@article{Engle2002,
  title = {Dynamic Conditional Correlation: A Simple Class of Multivariate Generalized Autoregressive Conditional Heteroskedasticity Models},
  author = {Engle, Robert},
  year = 2002,
  journal = {Journal of Business \& Economic Statistics},
  volume = {20},
  number = {3},
  doi = {10.1198/073500102288618487},
  pages = {339--350},
  publisher = {[American Statistical Association, Taylor \& Francis, Ltd.]},
  urldate = {2026-03-12}
}

@article{Engle2012,
  title = {Dynamic Equicorrelation},
  author = {Engle, Robert and Kelly, Bryan},
  year = 2012,
  month = apr,
  journal = {Journal of Business \& Economic Statistics},
  volume = {30},
  number = {2},
  pages = {212--228},
  publisher = {Taylor \& Francis},
  doi = {10.1080/07350015.2011.652048},
  urldate = {2026-03-15}
}

@article{Engle2019,
  title = {Large Dynamic Covariance Matrices},
  author = {Engle, Robert F. and Ledoit, Olivier and Wolf, Michael},
  year = 2019,
  month = apr,
  journal = {Journal of Business \& Economic Statistics},
  volume = {37},
  number = {2},
  pages = {363--375},
  publisher = {Taylor \& Francis},
  doi = {10.1080/07350015.2017.1345683},
  urldate = {2026-03-15}
}

@article{French1986,
  title = {Stock Return Variances: {{The}} Arrival of Information and the Reaction of Traders},
  shorttitle = {Stock Return Variances},
  author = {French, Kenneth R. and Roll, Richard},
  year = 1986,
  month = sep,
  journal = {Journal of Financial Economics},
  volume = {17},
  number = {1},
  pages = {5--26},
  doi = {10.1016/0304-405X(86)90004-8},
  urldate = {2026-03-22}
}

@article{Hansen2011,
  title = {The Model Confidence Set},
  author = {Hansen, Peter R. and Lunde, Asger and Nason, James M.},
  year = 2011,
  journal = {Econometrica},
  volume = {79},
  number = {2},
  pages = {453--497},
  doi = {10.3982/ECTA5771},
  urldate = {2026-03-12},
  copyright = {\copyright{} 2011 The Econometric Society},
  langid = {english}
}

@article{Hansen2026,
  title = {Convolution-{\emph{t}} Distributions},
  author = {Hansen, Peter Reinhard and Tong, Chen},
  year = 2026,
  month = mar,
  journal = {Journal of Econometrics},
  volume = {254},
  pages = {106212},
  doi = {10.1016/j.jeconom.2026.106212},
  urldate = {2026-03-07}
}

@article{Linton2020,
  title = {A Coupled Component {{DCS-EGARCH}} Model for Intraday and Overnight Volatility},
  author = {Linton, Oliver and Wu, Jianbin},
  year = 2020,
  month = jul,
  journal = {Journal of Econometrics},
  volume = {217},
  number = {1},
  pages = {176--201},
  doi = {10.1016/j.jeconom.2019.12.015},
  urldate = {2026-03-03},
  langid = {english}
}

@article{Lou2019,
  title = {A Tug of War: {{Overnight}} versus Intraday Expected Returns},
  shorttitle = {A Tug of War},
  author = {Lou, Dong and Polk, Christopher and Skouras, Spyros},
  year = 2019,
  month = oct,
  journal = {Journal of Financial Economics},
  volume = {134},
  number = {1},
  pages = {192--213},
  doi = {10.1016/j.jfineco.2019.03.011},
  urldate = {2026-03-03},
  langid = {english}
}

@article{Oh2023a,
  title = {Dynamic Factor Copula Models with Estimated Cluster Assignments},
  author = {Oh, Dong Hwan and Patton, Andrew J.},
  year = 2023,
  month = dec,
  journal = {Journal of Econometrics},
  volume = {237},
  number = {2},
  pages = {105374},
  doi = {10.1016/j.jeconom.2022.07.012},
  urldate = {2026-03-03},
  langid = {english}
}

@article{Tong2023,
  title = {Characterizing Correlation Matrices That Admit a Clustered Factor Representation},
  author = {Tong, Chen and Hansen, Peter Reinhard},
  year = 2023,
  month = dec,
  journal = {Economics Letters},
  volume = {233},
  pages = {111433},
  doi = {10.1016/j.econlet.2023.111433},
  urldate = {2026-03-03}
}

@article{Tong2026,
  title = {Cluster {{GARCH}}},
  author = {Tong, Chen and Hansen, Peter Reinhard and Archakov, Ilya},
  year = 2026,
  month = jan,
  journal = {Journal of Business \& Economic Statistics},
  volume = {44},
  number = {1},
  pages = {148--161},
  doi = {10.1080/07350015.2025.2510325},
  urldate = {2026-03-03},
  langid = {english}
}

@article{Tse2002,
  title = {A Multivariate Generalized Autoregressive Conditional Heteroscedasticity Model with Time-Varying Correlations},
  author = {Tse, Y. K. and Tsui, Albert K. C.},
  year = 2002,
  journal = {Journal of Business \& Economic Statistics},
  volume = {20},
  number = {3},
  pages = {351--362},
  doi={10.1198/073500102288618496},
  publisher = {[American Statistical Association, Taylor \& Francis, Ltd.]},
  urldate = {2026-03-12}
}

@article{TongHansen2026,
  author  = {Tong, Chen and Hansen, Peter Reinhard},
  title   = {Dynamic Factor Correlations},
  journal = {Journal of Applied Econometrics},
  year    = {2026},
   ISSN = {0883-7252
1099-1255},  
  note    = {Forthcoming},
  doi     = {10.1002/jae.70062},
     type = {Journal Article}
}

@article{Symitsi2018,
  title = {Covariance Forecasting in Equity Markets},
  author = {Symitsi, Efthymia and Symeonidis, Lazaros and Kourtis, Apostolos and Markellos, Raphael},
  year = 2018,
  month = nov,
  journal = {Journal of Banking \& Finance},
  volume = {96},
  pages = {153--168},
  doi = {10.1016/j.jbankfin.2018.08.013},
  urldate = {2026-06-26},
  langid = {english}
}

@article{Moura2020,
  title = {Comparing High-Dimensional Conditional Covariance Matrices: Implications for Portfolio Selection},
  author = {Moura, Guilherme V. and Santos, Andr{\'e} A. P. and Ruiz, Esther},
  year = 2020,
  month = sep,
  journal = {Journal of Banking \& Finance},
  volume = {118},
  pages = {105882},
  doi = {10.1016/j.jbankfin.2020.105882},
  urldate = {2026-06-26},
  langid = {english}
}

@article{DeNard2022,
  title = {Large Dynamic Covariance Matrices: Enhancements Based on Intraday Data},
  author = {De Nard, Gianluca and Engle, Robert F. and Ledoit, Olivier and Wolf, Michael},
  year = 2022,
  month = may,
  journal = {Journal of Banking \& Finance},
  volume = {138},
  pages = {106426},
  doi = {10.1016/j.jbankfin.2022.106426},
  urldate = {2026-06-26},
  langid = {english}
}

@article{PaolellaPolakWalker2021,
  title = {A Non-Elliptical Orthogonal GARCH Model for Portfolio Selection under Transaction Costs},
  author = {Paolella, Marc S. and Polak, Pawe{\l} and Walker, Patrick S.},
  year = 2021,
  month = apr,
  journal = {Journal of Banking \& Finance},
  volume = {125},
  pages = {106046},
  doi = {10.1016/j.jbankfin.2021.106046},
  urldate = {2026-06-26},
  langid = {english}
}

@article{Kim2024,
   author = {Kim, Donggyu and Oh, Minseog and Song, Xinyu and Wang, Yazhen},
   title = {Factor Overnight {GARCH-It\^{o}}  Models},
   journal = {Journal of Financial Econometrics},
   volume = {22},
   number = {5},
   pages = {1209-1235},
   DOI = {10.1093/jjfinec/nbad032},
   year = {2024},
   type = {Journal Article}
}

@article{Kim2023,
   author = {Kim, Donggyu and Shin, Minseok and Wang, Yazhen},
   title = {Overnight {GARCH-It\^{o}}  Volatility Models},
   journal = {Journal of Business \& Economic Statistics},
   volume = {41},
   number = {4},   
   pages = {1215-1227},
   doi = {10.1080/07350015.2022.2116027},
   year = {2023},
   type = {Journal Article}
}

@article{KangBabbs2012,
  author  = {Kang, Long and Babbs, Simon H.},
  title   = {Modeling Overnight and Daytime Returns Using a Multivariate Generalized Autoregressive Conditional Heteroskedasticity Copula Model},
  journal = {The Journal of Risk},
  year    = {2012},
  volume  = {14},
  number  = {4},
  pages   = {35--63},
  doi     = {10.21314/JOR.2012.246}
}

@article{HonigKircher2025,
  author  = {Honig, Igor and Kircher, Felix},
  title   = {Large Dynamic Covariance Matrices and Portfolio Selection with a
             Heterogeneous Autoregressive Model},
  journal = {Journal of Banking \& Finance},
  year    = {2025},
  volume  = {178},
  pages   = {107505},
  doi     = {10.1016/j.jbankfin.2025.107505}
}

@article{LockwoodMcInish1990,
  author  = {Lockwood, Larry J. and McInish, Thomas H.},
  title   = {Tests of Stability for Variances and Means of Overnight/Intraday
             Returns during Bull and Bear Markets},
  journal = {Journal of Banking \& Finance},
  year    = {1990},
  volume  = {14},
  number  = {6},
  pages   = {1243--1253}
}

@article{MoshirianNguyenPham2012,
  author  = {Moshirian, Fariborz and Nguyen, Huy G. L. and Pham, Peter K.},
  title   = {Overnight Public Information, Order Placement, and Price Discovery
             during the Pre-Opening Period},
  journal = {Journal of Banking \& Finance},
  year    = {2012},
  volume  = {36},
  number  = {10},
  pages   = {2837--2851},
  doi     = {10.1016/j.jbankfin.2012.06.007}
}

@article{GirardiErgun2013,
  author  = {Girardi, Giulio and Erg{\"u}n, A. Tolga},
  title   = {Systemic Risk Measurement: Multivariate {GARCH} Estimation of {CoVaR}},
  journal = {Journal of Banking \& Finance},
  year    = {2013},
  volume  = {37},
  number  = {8},
  pages   = {3169--3180},
  doi     = {10.1016/j.jbankfin.2013.02.027}
}

\end{document}